\documentclass[printer]{aa}
\usepackage{graphicx}
\usepackage[]{natbib}
\usepackage{amsmath}
\usepackage{amssymb}

\bibpunct{(}{)}{;}{a}{}{,} 

\def\kms{km~s$^{-1}$}
\def\tef{\textit{T}_{\text{eff}}}
\def\logg{\text{log}(\textit{g})}
\def\metah{[\text{M}/\text{H}]}
\def\feh{[\text{Fe}/\text{H}]}

\def\alffe{[\alpha/\text{Fe}]}
\def\snr{S/N}
\def\vsini{v\sin(i)}
\def\vrad{V_{\text{rad}}}
\def\loggf{\log{gf}}
\def\ccf{\text{FWHM}_{\text{CCF}}}
\def\ali{\text{A}_{\text{Li}}}
\def\cms{cm~s$^{-2}$}

\begin{document}

\title{The AMBRE Project: \\Constraining the lithium evolution in the Milky Way \thanks{\tablename{~\ref{catalogue}} is only available in electronic form at the CDS via anonymous ftp to cdsarc.u-strasbg.fr (130.79.128.5) or via http://cdsweb.u-strasbg.fr/cgi-bin/qcat?J/A+A/}}

\titlerunning{Lithium in FGK stars}
\authorrunning{G. Guiglion and collaborators}

\author{G. Guiglion \inst{1}, 
P. de Laverny \inst{1}, 
A. Recio-Blanco \inst{1}, 
C. C. Worley \inst{2}, 
M. De Pascale \inst{3}, 
T. Masseron \inst{2}, 
N. Prantzos \inst{4}, 
\v{S}.~Mikolaitis \inst{5}}

\institute{Universit\'e C\^ote d'Azur, Laboratoire Lagrange, Observatoire de la C\^ote d'Azur, CNRS, 
Blvd de l'Observatoire, CS 34229, 06304 Nice cedex 4, France 
\and Institute of Astronomy, University of Cambridge, Madingley Road, Cambridge CB3 0HA, United Kingdom 
\and INAF - Osservatorio Astronomico di Padova, Vicolo Osservatorio 5, 35122 Padova, Italy
\and Institut d'Astrophysique de Paris, UMR7095 CNRS, Universit\'e P. \& M. Curie, 98bis Bd. Arago, 75104 Paris, France 
\and Institute of Theoretical Physics and Astronomy, Vilnius University, Saul\.{e}tekio al. 3, LT-10222, Vilnius, Lithuania}

\date{Received 13/05/2016 ; accepted 16/07/2016}

\abstract{The chemical evolution of lithium in the Milky Way represents a major problem in modern 
astrophysics. Indeed, lithium is, on the one hand, easily destroyed in stellar interiors, and, on the other hand, 
produced at some specific stellar evolutionary stages that are still not well constrained.}
{The goal of this paper is to investigate the lithium stellar content of Milky Way stars in order to put 
constraints on the lithium chemical enrichment in our Galaxy, in particular in both the thin and thick discs.}
{Thanks to high-resolution spectra from the ESO archive and high quality atmospheric parameters, we were 
able to build a massive and homogeneous catalogue of lithium abundances for $7\,300$ stars derived with an automatic 
method coupling, a synthetic spectra grid, and a Gauss-Newton algorithm. We validated these lithium abundances
with literature values, including those of the Gaia benchmark stars.}
{In terms of lithium galactic evolution, we show that the 
interstellar lithium abundance increases with metallicity by $1\,$dex from $\metah=-1\,$dex to 
$+0.0\,$dex. Moreover, we find that this lithium ISM abundance decreases by about $0.5\,$dex 
at super-solar metalllicity. Based on a chemical separation, we also observed that the stellar lithium content 
in the thick disc increases rather slightly with metallicity, while the thin disc shows a steeper increase. 
The lithium abundance distribution of $\alpha$-rich, metal-rich stars has a peak at $\ali\sim3\,$dex.}
{We conclude that the thick disc stars suffered of a low lithium chemical enrichment, showing lithium abundances 
rather close to the Spite plateau while the thin disc stars clearly show an increasing lithium chemical enrichment 
with the metallicity, probably thanks to the contribution of low-mass stars.}


\maketitle

\section{Introduction}

Nowadays, the lithium chemical element represents a major centre of interest because 
its chemical evolution history in stars and in the Milky Way is still uncertain. 
On the one hand, according to the Standard Big Bang 
Nucleosynthesis model (SBBN), the primordial lithium abundance is predicted to be 
$\ali\sim2.6$ \citep{spergel_2003}; however, this primordial lithium abundance is in 
strong disagreement with the photospheric lithium abundance measured in old metal-poor 
dwarfs ($\ali=2.2\,$dex) known as the Spite plateau \citep{spite_1982}. 
On the other hand, the lithium meteoritic abundance is known to be $\ali=3.26\,$dex \citep{lodders_2009} 
indicating a certain enrichment since the Big Bang,  whereas the solar lithium photospheric 
abundance is sensitively lower, i.e. $\ali=1.05\,$dex, \citep{grevesse_2007} highlighting internal destruction. Indeed, 
lithium is easily destroyed in stellar interiors ($T\sim2\times10^6\,$K) by (p, $\alpha$) reactions and, because of mixing events, lithium can be strongly depleted at the stellar surface.\\

Lithium can also be produced in the interstellar medium (ISM) via spallation 
by Galactic cosmic rays (GCC; \citealt{reeves_1970}) and at very specific phases 
of the stellar evolution: core-collapse supernovae (CCSN; 
\citealt{hartmann_1999}); novae \citep{arnould_1975}, as supported by recent observation of 
\citep{2015Natur.518..381T}; and  low-mass giants via \emph{cool bottom burning} 
\citep{sackmann_1999}, as observed first by \citet{wallerstein_1982}. Finally, 
asymptotic giant branch stars (AGB) can produce lithium via \emph{hot bottom burning} 
\citep{sackmann_1992, abia_1999} as first observed by \citet{mc_kellar}.

Nevertheless, even if several production sites are known, these stellar yields are still 
not well constrained. For example, adopting
yields from \citet{woosley_1990}, \citet{travaglio_2001} 
estimated that CCSN could produce $40\%$ of the meteoritic abundance, while their contribution falls down to $10\%$ when adopting 
hydrodynamical models \citep{heger_2000}. Building chemical evolution 
models of lithium in the Milky Way is thus not an easy task \citep{romano_2001, prantzos_2012}. 
Moreover, when comparing observations to chemical evolution models, the lack of good age estimates does 
not allow us to study the lithium evolution with stellar age \citep{lambert_2004, delgado_2015}.

In order to investigate the lithium evolution in the Milky Way, one needs a statistically 
robust and homogeneous sample, such that a large metallicity domain can be covered. Up to now, 
very few studies presented small or inhomogeneous samples of few hundreds of stars \citep{lambert_2004, 
ramirez_2012, delgado_2015}. To provide more robust information on the lithium evolution in the 
Galaxy, we therefore performed a chemical study of the lithium behaviour from a homogeneous 
and very large catalogue of abundances created from the ESO archive without 
any external addition of smaller catalogues. This study is placed in the context of the AMBRE 
Project \citep{delaverny_2013}. The automatic determination of lithium has been performed for 
$7\,300$ non-rotating and assumed non-binary stars, covering a wide range of metallicity. 

The paper is organized as follows: in Sect~\ref{observationnal_data} we present the 
spectroscopic data used for our analysis while in Sect~\ref{metohd} we detail our 
automatic method of lithium derivation. The AMBRE catalogue of lithium abundances 
is presented in Sect~\ref{catalouge}. We validate our lithium measurements in 
Sect~\ref{compare_extern}. Finally, the lithium evolution in the Galaxy is discussed 
in Sect~\ref{discussion} in the context of recent chemical evolution models of the Milky 
Way taking the thin to thick dichotomy into account. We conclude this work in 
Sect~\ref{conclusiooooonnnn}.

\section{Observational data set from the AMBRE project}\label{observationnal_data}

This lithium study is based on spectroscopic data from the AMBRE project.
The goal of this project is to parametrize the HARPS, FEROS, UVES, and GIRAFFE spectral archives 
\citep{delaverny_2013}, providing robust automatic determinations of the radial velocity ($\vrad$), effective 
temperature ($\tef$), surface gravity ($\logg$), metallicity ($\metah$), and global $\alpha$ enrichment 
with respect to iron ($\alffe$) together with their associated errors. In the present paper, we 
work with the first three samples that have already been parametrized: $3\,628$ UVES spectra in the Red580 set-up 
containing the Li feature at $6708\,$\AA~\citep{2016arXiv160208478W}, 
$89\,183$ HARPS spectra \citep{depascale_2014}, and $5\,981$ FEROS spectra \citep{worley_2012}. These 
analysed samples were built from the spectra corresponding to stars with $+1\le\logg\le+5\,$\cms~and a quality 
flag lower or equal to 1 (see e.g. \citealt{worley_2012} for details on this label). 
The HARPS sample actually consists of repeated spectra for most of stars, thus the number of 
analysed stars is much lower. The typical errors on $\tef$, $\logg$ and $\metah$ are 
[$108\,$K, $0.16\,$\cms, $0.10\,$dex] for UVES, [$93\,$K, $0.26\,$\cms, $0.08\,$dex] for HARPS, and 
[$120\,$K, $0.20\,$\cms, $0.10\,$dex] for FEROS. In the following, we also use the AMBRE estimates 
of the signal-to-noise ratio ($\snr$) and the FWHM of the cross-correlation function used to 
derive $\vrad$ for a given star ($\ccf$).

\section{Automatic lithium abundance analysis}\label{metohd}

The lithium abundances of the AMBRE spectra were 
automatically derived via an optimization method, 
by coupling a pre-computed synthetic spectra grid 
and the GAUGUIN Gauss-Newton algorithm \citep{gauguin}. 
We detail here the steps of our procedure.

\subsection{The high-resolution synthetic spectra grid around the Li doublet}\label{gride}

As the AMBRE stellar sample covers a wide range of atmospheric 
parameters, a careful examination of atomic and molecular contributions 
to the emerging spectrum was performed to build the line list for the 
synthetic spectra grid computation. 

We first started with an atomic 
line list over $15\,$\AA~from the Vienna Atomic Line Database (VALD3; 
\citealt{kupka_1999, kupka_2000}). 
The line lists of nine molecular species were also taken into account:
CN \citep{Sneden2014}, TiO (Plez, priv. comm.), C2 \citep{Brooke2013, Ram2014}, 
CH \citep{2014A&A...571A..47M}, ZrO (Plez, priv. comm.), OH (Masseron, priv. comm), CaH 
(Plez, priv. comm.), VO (Plez, priv. comm.), and SiH \citep{1992RMxAA..23...45K}. 
We focused on a small wavelength 
range around the ${}^7\text{Li}$ doublet (located at $6707.8\,$\AA), 
from $6\,707.0$ to $6\,708.5\,$\AA, at $R=40\,000$. This narrow domain 
is sufficiently wide to measure large lithium features with our automatic 
procedure of lithium derivation. For the ${}^7\text{Li}$ doublet at 
$6\,708\,$\AA, we adopted the hyper-fine structure, consisting of six 
components. The wavelengths were taken from \citet{sansonetti_1995}, while 
the oscillator strengths come from the calculations of \citet{yan_1998}.

Our final line list was carefully calibrated with the Sun and Arcturus. We used the solar spectrum from 
\citet{neckel_1999} and the Hinkle Arcturus Atlas \citep{hinkle_2003}. For Arcturus, 
we considered the atmospheric parameters from \citet{ramirez_2011}. 
We only calibrated the main atomic lines contributing to the opacity (see 
\tablename{~\ref{raies_flux_vald_final_calib}}). We increased the 
oscillator strength for the $\ion{V}{I}$ line at $\lambda=6\,707.518\,$\AA~compared to previous 
studies (about $1.1\,$dex with respect to \citealt{ghezzi_2009}) to reproduce the red part of the blend in Arcturus, as this $\ion{V}{I}$ line has a small 
contribution in the Sun. As another source of opacity is clearly missing in the Sun at 
$\lambda\approx6\,708.0\,$\AA, 
we included an additional line as already assumed in several previous works. 
\citet{muller_1975} first proposed adding a $\ion{Si}{I}$ line with $\chi_e=6.00\,$eV 
to match the solar spectrum while later \citet{mandell_2004} independently tested
the two lines $\ion{Ti}{I}$ and $\ion{Ti}{II}$. We rejected the Ti contribution that clearly 
overestimates the strength of the feature in Arcturus, whereas the $\ion{Si}{I}$ line provided 
a satisfying fit for both reference stars. The final fit between the observed solar and Arcturus 
spectra were very good with a typical flux discrepancy equal to 0.07 and $0.23\,$\% per pixel, respectively.

\begin{table}
\centering
\begin{tabular}[c]{c c c c}
\hline
\hline 
Element  & Wavelength & $\chi_e$ & $\loggf$    \\
         &   (\AA)    &   (eV)   &                 \\
\hline
\hline
$\ion{Fe}{I}$      & 6707.070  & 5.273    & \textbf{-2.480} \\
$\ion{Fe}{I}$      & 6707.172  & 5.538    & \textbf{-2.600} \\
$\ion{Fe}{I}$      & 6707.431  & 4.608    & \textbf{-2.175} \\
$\ion{Sm}{II}$     & 6707.473  & 0.933    & -1.910          \\
$\ion{V}{I}$       & 6707.518  & 2.743    & \textbf{-0.800} \\
$\ion{Cr}{I}$      & 6707.596  & 4.208    & -2.625          \\
$\ion{Li}{I}$      & 6707.756  & 0.000    & -0.428          \\
$\ion{Li}{I}$      & 6707.768  & 0.000    & -0.206          \\
$\ion{Li}{I}$      & 6707.907  & 0.000    & -1.509          \\
$\ion{Li}{I}$      & 6707.908  & 0.000    & -0.807          \\
$\ion{Li}{I}$      & 6707.919  & 0.000    & -0.807          \\
$\ion{Li}{I}$      & 6707.920  & 0.000    & -0.807          \\
$\ion{Si}{I}$      & 6708.023  & 6.000    & \textbf{-2.820} \\
$\ion{V}{I}$       & 6708.094  & 1.218    & \textbf{-2.810} \\
$\ion{Ce}{II}$     & 6708.099  & 0.701    & -2.120          \\
$\ion{Fe}{I}$      & 6708.282  & 4.988    & \textbf{-2.630} \\
$\ion{Fe}{I}$      & 6708.347  & 5.486    & -2.506          \\
\hline
\end{tabular}
\caption{Main atomic lines around the ${}^7\text{Li}$ doublet. 
Astrophysically calibrated $\loggf$ values are highlighted in boldface.}
\label{raies_flux_vald_final_calib}
\end{table}

Based on this line list, a specific synthetic spectra grid was computed using the MARCS atmosphere models 
\citep{gustafsson_2008} and the LTE TURBOSPECTRUM code \citep{plez_2012}. 
Five dimensions were considered for this grid: $\tef$, $\logg$, $\metah$, $\alffe,$ and $\ali$
\footnote{$\ali=\log{\epsilon(\text{Li})}$; both notations are in 
logarithmic scale of number of atoms where $\log{\epsilon(\text{H})}=\text{A}_{\text{H}}=12$.}. 
The ranges of the atmospheric parameters are those 
of the AMBRE grid, $3\,000\le\tef\le8\,000\,$K (in steps of $200\,$K below 
$4000\,K$ and $250\,$K above), $+0\le\logg\le+5.5\,$\cms (in steps of $0.5\,$\cms), 
$-5\le\metah\le+1\,$dex (see Fig.2 in \citet{laverny_2012} for more details on the steps in 
$\metah$), whereas the variation in $\ali$ ranges from $-1$ to $+4\,$dex 
with a step of $+0.2\,$dex (26 different values of $\ali$).

The microturbulence velocity ($\xi$) was included in the grid computation by 
adopting $\xi$ varying as a function of $\tef$, $\logg,$ and $\feh$ as adopted in the Gaia-ESO Survey 
(Bergemann et al., in preparation, based on $\xi$ determinations from literature samples). 
The total number of synthetic spectra is $358\,335$, 
computed on a wavelength range of $15\,$\AA, centred on the lithium doublet at 
$6\,708\,$\AA\  with a sampling of $0.004\,$\AA~and a spectral resolution 
thatis higher than $150\,000$. 

\subsection{Preparing the set of synthetic and observed spectra}\label{prep}

One of the main characteristics of our method is that we do not synthesize on the fly model spectra to
fit the observed spectrum. We interpolate the pre-computed 
5-D synthetic spectra grid of Sect.~\ref{gride} at the atmospheric parameters of 
the target derived within the AMBRE project to prepare a small 
set of interpolated synthetic spectra for a direct comparison 
with the observation. These spectra are interpolated in a first step. This interpolating spectra 
procedure allows us to derive abundances very quickly for large sets of observations.

More precisely, we build a set of interpolated synthetic spectra, composing a 1-D grid $S(\ali, \lambda)$ 
on the lithium abundance. Practically speaking, a cubic interpolation of the 5-D synthetic spectra grid 
was performed at the four atmospheric parameters of the observed star $\tef^{\star}$, $\logg^{\star}$, $\feh^{\star}$, 
and $\alffe^{\star}$, resulting in the 1-D grid $S(\ali, \lambda)$. As the variation in flux can be non-linear, 
we performed a Catmull-Rom interpolation based on 
B\'ezier curves. For a given lithium abundance of the 1-D grid, the corresponding synthetic interpolated spectrum 
is a combination of $4^4=256$ synthetic spectra because the four closest spectra in each parameter dimension 
($\tef$, $\logg$, $\metah$, and $\alffe$) are considered and weighted.

For stars with parameters close to the edges of the synthetic spectra grid, the 256 spectra are not 
systematically found and a simple linear interpolation is carried out. Roughly half of the AMBRE stars 
are interpolated linearly. The resulting 1-D grid $S(\ali, \lambda)$ in lithium abundance at $\tef^{\star}$, 
$\logg^{\star}$, $\feh^{\star}$, and $\alffe^{\star}$ varies from $-1$ to $+4\,$dex and is composed of 
26 model spectra. 

Concerning the observed spectra $O(\lambda)$, their resolution was degraded to $40\,000$ considering 
an instrumental Gaussian profile. This value was chosen as being the lowest spectral resolution 
of the three spectrographs FEROS, UVES, and HARPS. This allows us to analyse all the AMBRE spectra in a 
very homogeneous way. Respecting the Shannon theorem, we re-sampled these spectra 
to a pixel size of $0.05\,$\AA. We applied the same convolution and re-sampling to 
the synthetic spectra grid. Then, to put the spectra in the same rest frame, we performed 
the radial velocity correction with the value provided by the AMBRE project.

Finally, an automatic adjustment of the continuum was performed on the observed spectrum. 
For that purpose, we adopted an interpolated spectrum at the atmospheric parameters of the star. 
We removed the line features by $\sigma$-clipping, and a ratio between the synthetic flux and the 
observed flux is computed over a spectral range of $15\,$\AA, centred on the lithium doublet. 
This ratio is consecutively fitted by a third order polynomial function, $\sigma$-clipped, and then fitted 
a second time. Finally, the observed spectra are divided by the polynomial fit to adjust 
their continuum. This step was validated automatically by checking that the normalization is consistent over a much 
broader spectral range of $100\,$\AA. Systematic errors on $\ali$ determination due to automatic continuum 
placement were not estimated, but should be negligible, for example compared  to the errors of the effective temperature and 
taking into account that our sample is mainly composed of high $\snr$ spectra.

\begin{figure}
\centering
\includegraphics[width=1.0\linewidth]{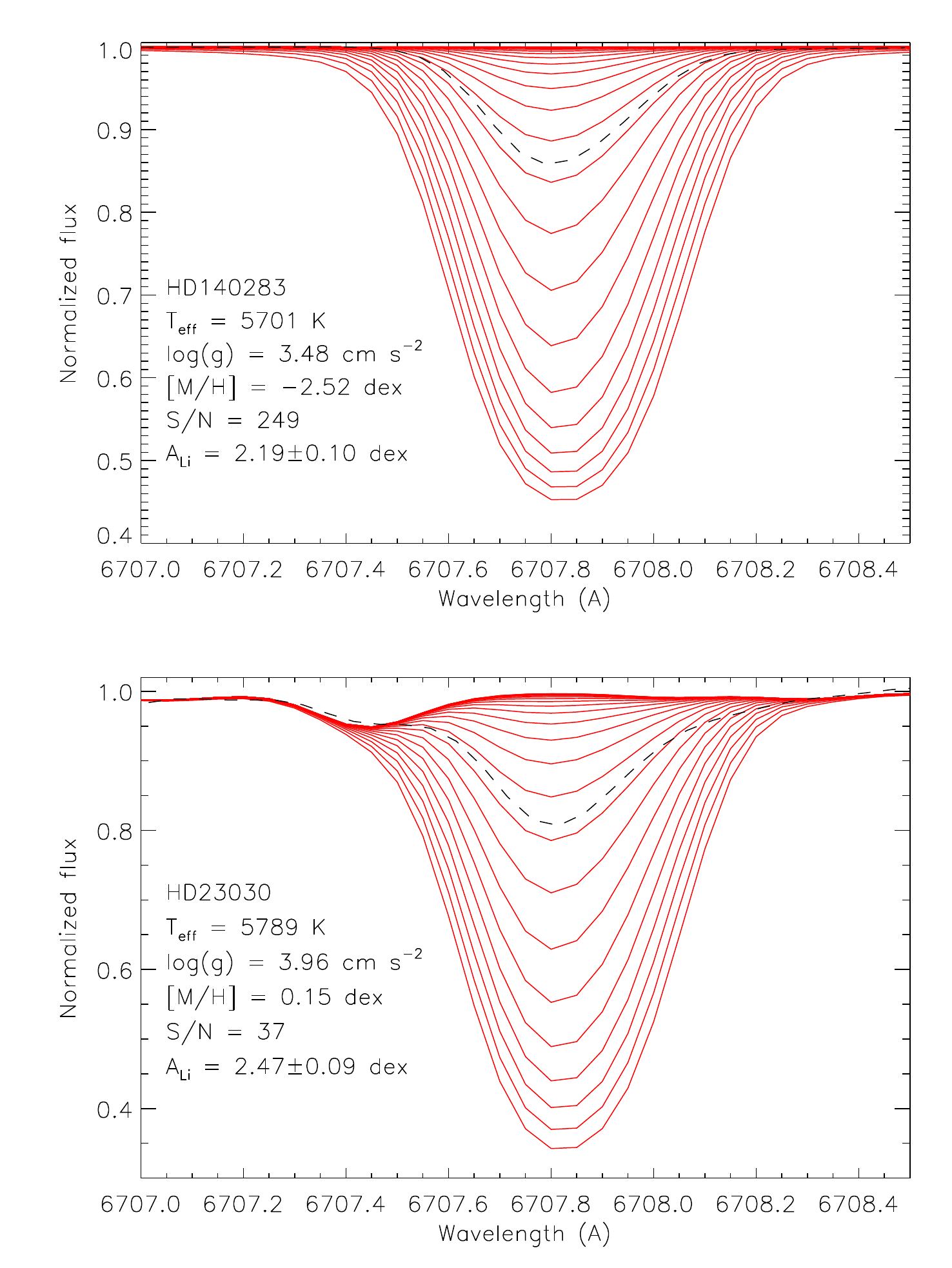}
\caption{\label{plot_grideee}Observed spectrum (dashed line) of HD140283 (top) and HD23030 (bottom) at $R=40\,000$. 
The corresponding 1-D synthetic spectra grid with lithium variations from -1 to $4\,$dex and a step of $0.2\,$dex, 
with the grid interpolated at the atmospheric parameters of both stars, is shown in red.}
\end{figure}

We show two typical examples of 1-D grids and observed spectra in \figurename~\ref{plot_grideee}, for 
HD140283 and HD23030, observed with a $\snr$ of 249 and 37, respectively.

\subsection{Derivation of the lithium abundances with GAUGUIN}

From the 1-D lithium grid $S(\ali, \lambda)$ described above, we compute a quadratic distance 
$D(\ali, \lambda)$ between the observed spectrum $O(\lambda)$ and each point of the grid $S(\ali, \lambda)$ 
over a wavelength range of $1.5\,$\AA~centred on the lithium doublet.
The minimum of $D(\ali, \lambda)$ provides a first guess $\ali^0$ of the solution, \textit{i.e.} 
the closest point of the grid. 
Then, this first guess is optimized via the Gauss-Newton algorithm GAUGUIN \citep{gauguin}. 
For that purpose, from $S(\ali, \lambda)$, a synthetic spectrum is interpolated at $\ali^0$ and 
the correction computed by GAUGUIN is then 
\begin{equation}\label{eq_gaug}
\delta_{\ali} = (\mathbf{J}^T\mathbf{J})^{-1}\mathbf{J}^T[O(\lambda) - S(\ali^0, \lambda)],
\end{equation}
where $\mathbf{J}$ is the Jacobian matrix $[\partial S(\ali^0, \lambda)) / \partial\ali]$. 
As the curve of growth is highly non-linear with the abundance, the flux in the lithium line varies non-linearly 
as well and the Catmull-Rom interpolation is again used to build the interpolated flux and derivatives. 
The algorithm stops when the distance is minimal between the interpolated spectrum $S(\ali, \lambda)$ and 
$O(\lambda)$. A wavelength range (large enough to safely measure the highest abundances) of $1.5\,$\AA~centred 
on the lithium doublet is considered for this operation. Upper limits are provided when the lithium 
feature is too weak with respect to the $\snr$ of the spectrum. 
On the other hand, we can detect lower limits for $\ali\ge4\,$dex.

The simple application of GAUGUIN, which is basically lower than 10 msec, is very fast for a single spectral line.
This completely automatic procedure (including all the synthetic spectra preparation 
described in Sect.~\ref{prep}) is fast and adapted for massive spectroscopic surveys. Basically, a single 
Li line abundance can be derived in $\approx1.5$ seconds. The pipeline was implemented combining IDL and C++ 
languages. 

The errors on $\ali$ were estimated by propagating the errors on the three atmospheric parameters 
$\big\{\tef^{\star},\,\logg^{\star},\,\metah^{\star}\big\}$ provided by AMBRE and summing them quadratically (see 
Sect.~\ref{catalouge} and \figurename{~\ref{eali_teff}}). The error on the lithium measurement 
is dominated by the error on the effective temperature. 
Furthermore, we checked that a typical error of $1\,$\kms~on the radial velocity leads to 
a negligible error on $\ali$. 
Also, by generating Sun and Arcturus synthetic spectra for different values of $\xi$ and $\ali$ 
from 1 to $3\,$dex ($R=40\,000$), we checked the impact of a pessimistic error on 
$\xi$ of $1\,$\kms~for the dwarfs and 
$2\,$\kms~for the giants. For the Sun, the error defined by $e_{\ali}/\ali$ is largely below $1\%$ 
whatever the value of $\ali$ and can thus be neglected compared to the error contribution of the 
atmospheric parameters. For Arcturus, the error is also low ($4\%$ at $2\,$dex and $5\%$ at 
$3\,$dex), which is well below the typical errors caused by the other atmospheric parameters for the 
AMBRE giants. This $\xi$ error contribution has thus been neglected in the derivation of the total error.

Finally, we assume, for this lithium analysis, that all the targets are single stars 
since binary detection is not a part of the AMBRE parametrization pipeline.

\subsection{Non-LTE corrections of the AMBRE/Li abundances}\label{nlte_corr}
As the AMBRE stellar sample and synthetic spectra grid cover a wide range of atmospheric parameters, 
it is important to take possible non-LTE effects on the AMBRE/Li abundances  into account. For 
that purpose, we adopted the non-LTE corrections presented by \citet{lind_2009} to estimate the corrections 
that we have to apply to our Li measurements. For these NLTE corrections, we assumed 
$\metah=+0\,$dex for stars with $\metah>0\,$dex because the dependence with $\metah$ is low (K. Lind, private communication). 
Moreover, we adopted the NLTE corrections at $\metah=-3\,$dex for the few metal-poor stars with $\metah<-3\,$dex, as 
adopted for example by \citet{sbordone_2010}. We also did not publish 
the NLTE corrections for stars outside of the Lind et al. ranges ($~2\%$ of the AMBRE catalogue), mostly when $\tef<4000\,$K. 
Finally, for the most lithium-poor stars with $\ali$ lower than 
the lowest value of the Lind et al. LTE curve-of-growth ($-0.3\,$dex, variable with $\tef$), no corrections were computed; 
this subsample represents $~9\%$ of the AMBRE catalogue. The resulting corrections are presented in Sect.~\ref{catalouge} (see 
\tablename{~\ref{catalogue}}, column 7) for $89\%$ of the AMBRE catalogue.

\section{The AMBRE catalogue of lithium abundances}\label{catalouge}

The AMBRE/Li catalogue is presented in \tablename{~\ref{catalogue}. We 
first looked for repeated observations separately in each of the UVES, HARPS, and FEROS samples. For UVES and FEROS, we performed a cross-match on the spectra coordinates 
with a radius of 2 and $10\,$arcsec on the sky, leading to remaining samples composed of 
$1\,031$, and $3\,526$ stars, respectively. For a given star with several 
spectra collected with the same spectrograph, we chose the spectrum with the best 
atmospheric parameters in terms of the AMBRE $\chi^2$ quality flag (see 
\citealt{worley_2012}), leading to the best lithium measurement}. For HARPS, as the number of repeated spectra for a given 
star can be huge with different target stars very close to each other, we adopted the sample of 
$4\,355$ stars from Mikolaitis et al. 2016 (in preparation) 
based on a search of both coordinates and atmospheric parameters differences. The last step to merge 
these UVES, HARPS, and FEROS samples together was to identify stars possibly observed with two or 
three spectrographs with a new coordinate cross-match on a radius of $10\,$arcsec.  In this case, 
for a given star, we selected its spectrum with the best parameters in terms of $\chi^2$ as explained above. 
The final working AMBRE catalogue is then composed of $7\,821$ stars, 
subdivided into $3\,301$ FEROS, $878$ UVES, and $3\,642$ HARPS stars.

We are aware that the synthetic spectra grid was computed with no rotation, assuming 
that the stars are slow rotators. In order to estimate the impact of the rotation on the derived 
AMBRE/Li abundances, we artificially broadened solar and Arcturus synthetic spectra for 
different $\ali$ values. Using typical $\vsini$ values lower than $10\,$\kms~for Arcturus and 
$15\,$\kms~for the Sun, the estimated errors on $\ali$ are found to be lower than errors due to 
the atmospheric parameters. We therefore adopted these $\vsini$ values as the largest acceptable values 
to derive $\ali$. Thus, since AMBRE does not provide the rotational velocity and
only provides the FWHM of the cross-correlation function ($\ccf$), we looked for a relation between 
$\vsini$ and $\ccf$ for the AMBRE spectra. First, based on rotational velocities for stars from 
the FEROS sample \citep{laverny_2013}, we searched for a limit in terms of $\ccf$ (that is 
sensitively correlated to $\vsini$) that was valid over the whole atmospheric parameter range. 
We established a limit at $\ccf=20\,$\kms~for FEROS targets. We thus removed stars with 
higher $\ccf$ ($320$ stars, $10\%$ of the FEROS sample). 
Concerning the HARPS targets, showing lower $\ccf$ because of the larger resolution, 
we applied the same strategy as 
for FEROS, but based on $\vsini$ estimations from Mikolaitis et al. 2016 (in preparation). 
The established limit is $\ccf=15\,$\kms~(removing then 
$150$ stars, $4\%$ of the HARPS sample). For UVES stars, 
the $\ccf$ distribution shows intermediate values between HARPS and FEROS. As we do not 
have any $\vsini$ determinations for these stars, we applied a very pessimistic cut at $\ccf=15\,$\kms~
(removing then $79$ stars, $9\%$ of the UVES sample), 
as that adopted for the higher HARPS spectra resolution. For all these rotating stars, 
we decided not to publish their $\ali$ abundance.

As a consequence, the AMBRE/Li catalogue finally contains $7\,272$ stars
\footnote{Stars with close coordinates but different atmospheric 
parameters were clearly tagged by adopting a different name in the AMBRE/Li catalogue.}, 
\emph{i.e.} $\sim93\%$ of the stars for which we derived $\ali$, subdivided into $2\,981$ 
FEROS , $799$ UVES, and $3\,492$ HARPS stars. Among these, 
$6\,479$ stars have NLTE corrections. Up to now, this catalogue is the largest 
ever created, and is more than one order of magnitude larger than previous studies 
(\citet{delgado_mena_2015, delgado_2015}, 353 and 326 dwarf stars, respectively; 
\citet{ramirez_2012}, 671 dwarf stars; \citet{liu_2014} 378 giant stars). Moreover, 
this AMBRE/Li catalogue is perfectly homogeneous in terms of atmospheric 
parameters and $\ali$ determinations. The lithium measurements 
and NLTE corrections of these $7\,272$ stars are presented in \tablename{~\ref{catalogue}}. 
Atmospheric parameters of the AMBRE/Li stars are available in \citet{worley_2012, 2016arXiv160208478W} 
and in \citet{depascale_2014}.

\begin{figure}
\centering
\includegraphics[width=1.0\linewidth]{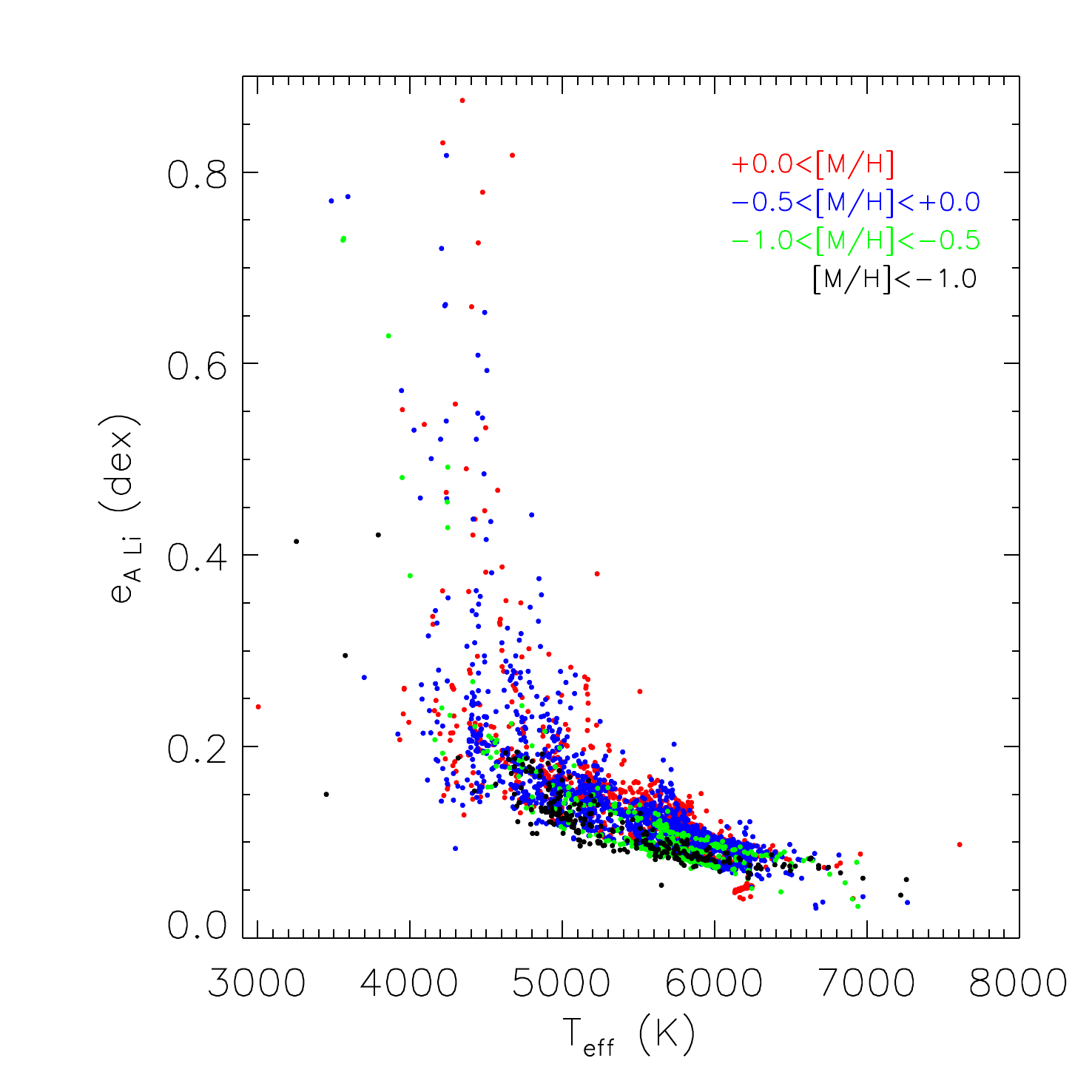}
\caption{\label{eali_teff} Errors on $\ali^{\text{ETL}}$ of the AMBRE 
catalogue of lithium measurements (excluding upper limits), colour-coded in bins of $\metah$.}
\end{figure}

Together with $\ali$, we also provide the errors on $\ali$ in \tablename{~\ref{catalogue}}. 
We show these errors on $\ali$ in \figurename{~\ref{eali_teff}}. We see that the error strongly 
increases for lower effective temperatures and tend to be lower for metal-poor 
stars for which the determinations are more robust (less blend). 
Almost $98\%$ of the stars have an error that is lower than $0.3\,$dex, while $40\%$ of 
the sample (the hottest one with $\tef>5\,500\,$K) have an error lower than $0.1\,$dex, which is low enough 
for conducting a detailed scientific application of our lithium measurements.

\begin{figure*}
\centering
\includegraphics[width=0.8\linewidth]{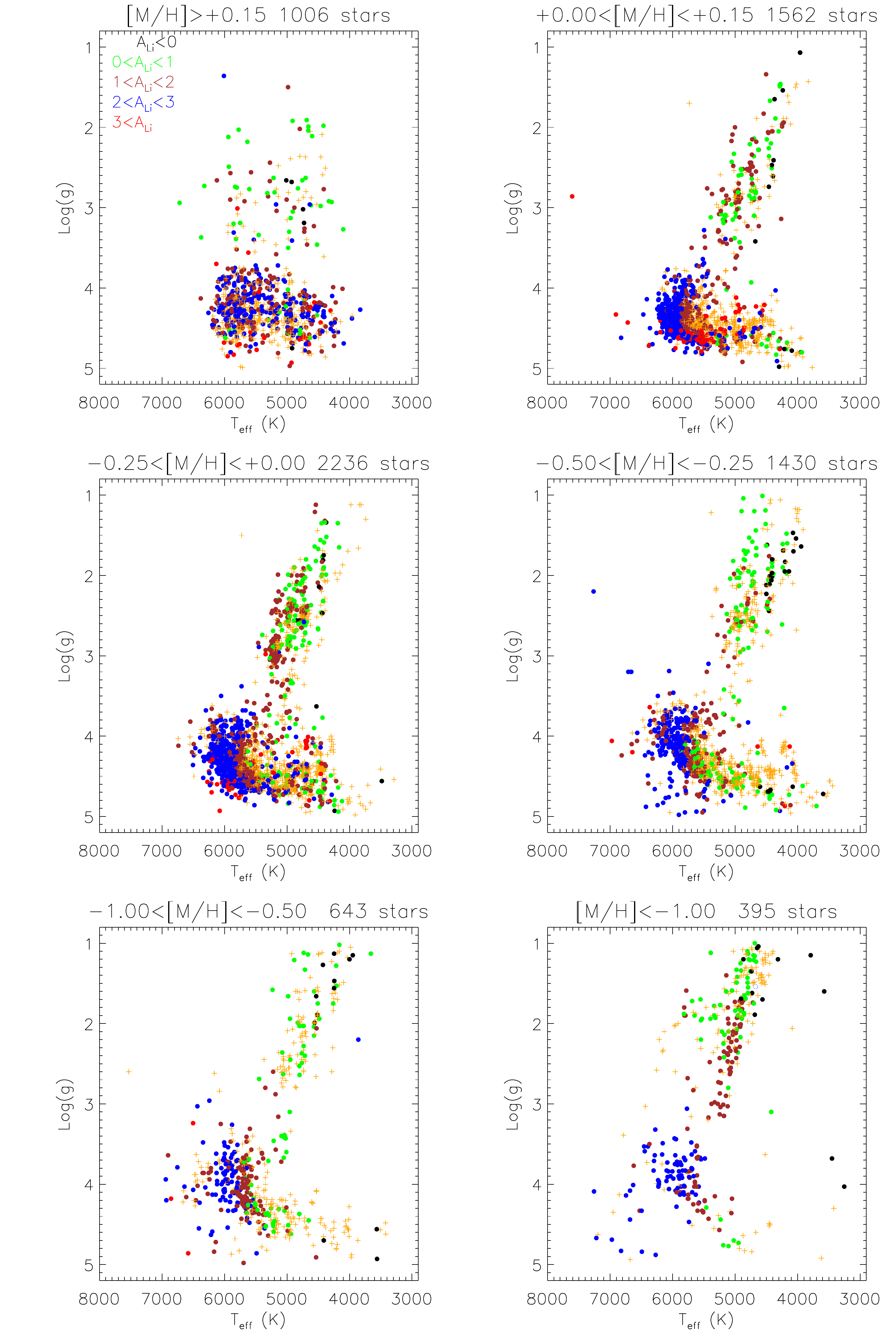}
\caption{\label{catalogue_hr} $\tef\,vs.\,\logg$ for the AMBRE catalogue of 
lithium measurements colour-coded in bins of $\ali$ (LTE). Filled and open circles correspond 
to lithium detections and $\ali>4.2\,$dex, respectively. Upper limits are represented by orange "+" symbol.}
\end{figure*}

We also show in \figurename{~\ref{catalogue_hr}} the Hertzsprung-Russel diagram of this AMBRE 
catalogue of lithium abundances for slow rotators. Thanks to the binning in $\metah$, 
we clearly see that the statistics in each stellar population are high, from metal-poor 
to metal-rich stars.

\begin{figure}
\centering
\includegraphics[width=1.0\linewidth]{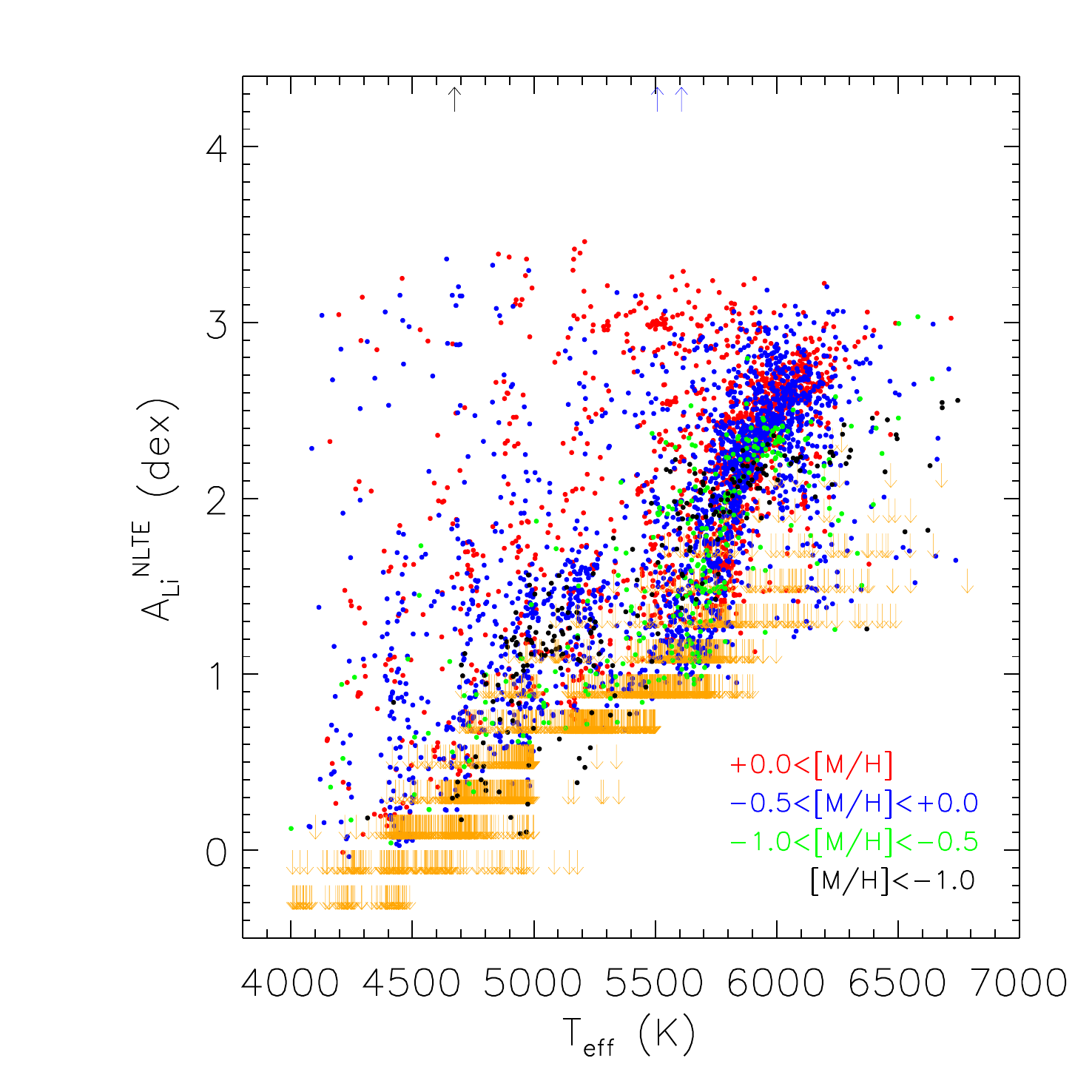}
\caption{\label{ali_teff} $\ali^{\text{NLTE}}\,vs.\,\tef$ with a colour-code in $\metah$. 
Upper and lower limits are symbolized by downwards and upwards arrows, respectively.} 
\end{figure}

We present the behaviour of $\ali^{\text{NLTE}}$ with $\tef$ in 
\figurename{~\ref{ali_teff}}. We see that cooler stars exhibit lower lithium abundances 
probably because of their deeper convective zone leading to higher destruction of lithium (
$\sim80\%\,$ of the AMBRE sample is composed by dwarf stars). We also observe a 
strong correlation between the lower limits of lithium with the effective temperature. We can 
understand this trend by considering that for a given $\snr$ value, the detectable lithium feature 
is stronger at lower temperatures.

Finally, we detected the presence of lithium-rich 
giants in the AMBRE/Li catalogue. These stars will be studied in a work in preparation.

\begin{table}
\centering
\begin{tabular}[c]{l c c c}
TARGNAME     &  Spectro        & $\text{A}_{\text{Li}}^{\text{LTE}}$ & $\Delta_{\text{NETL}}$ \\
            &                  & (dex)                               & (dex) \\
\hline
\hline
HD162396 & H & $+2.32\pm0.08$ &-0.01 \\
HD199288 & H & $+0.97\pm0.09$ &+0.01 \\
HD90422 & H & $+1.61\pm0.08$ &-0.01 \\
HD128167 & H & $<+1.40$ &- \\
HD091324 & H & $+1.95\pm0.07$ &-0.01 \\
HD215257 & H & $+2.27\pm0.08$ &-0.03 \\
HD211998 & H & $+1.20\pm0.10$ &+0.03 \\
... & ... & ... & ... \\
\hline
\end{tabular}
\caption{\label{catalogue}Identifier, spectrograph (U=UVES, H=HARPS, F=FEROS), 
lithium abundances, and NLTE-corrections (defined by $\Delta_{\text{NETL}} =\text{A}_{\text{Li}}^{\text{NLTE}} - 
\text{A}_{\text{Li}}^{\text{LTE}}$) of the AMBRE catalogue 
of lithium measurements.}\label{ambre_cata}
\end{table}

\section{Internal \& external validation}\label{compare_extern}

\subsection{Internal comparison between HARPS and UVES targets}

\begin{figure}
\centering
\includegraphics[width=1.0\linewidth]{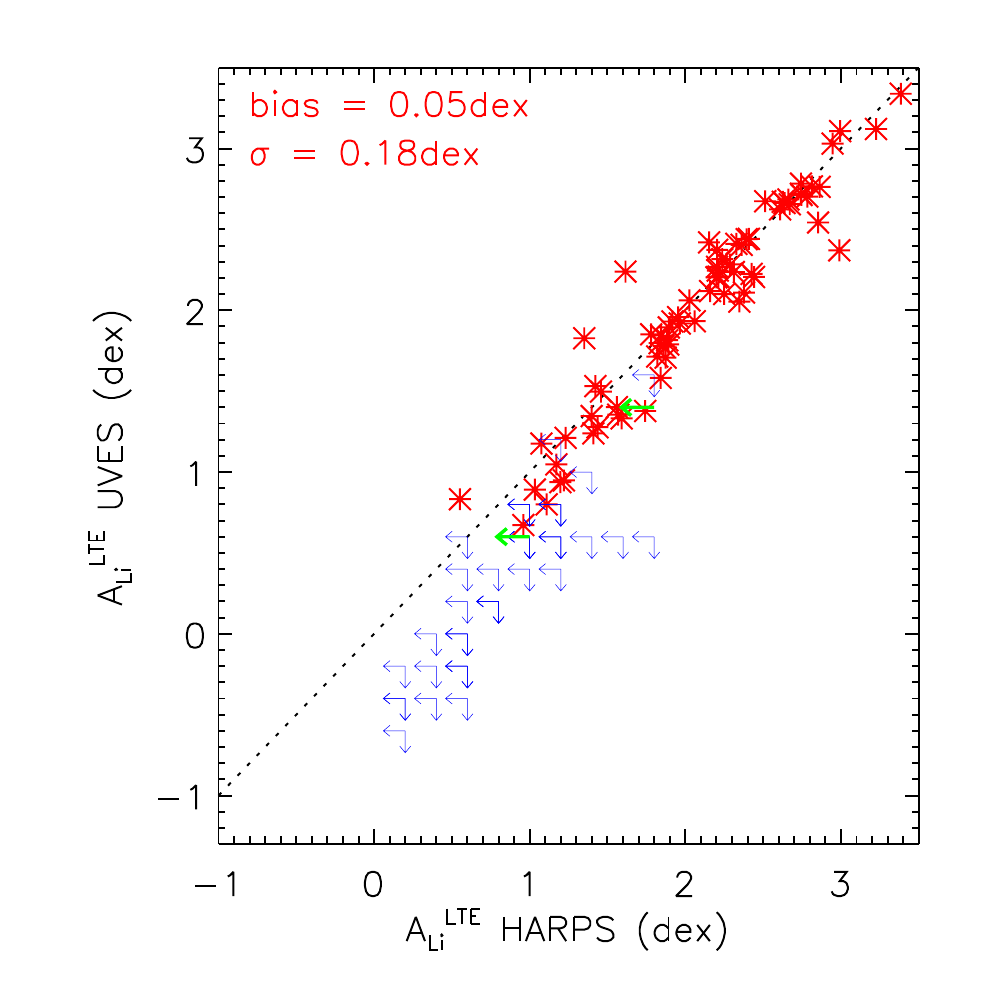}
\caption{\label{compare_harps_uves} Lithium abundances for a set 
of 117 dwarf and giant AMBRE stars observed both with UVES and HARPS. Lithium detections 
are shown in red, while upper limits are represented by blue and green arrows. Mean difference and standard 
deviation for the lithium detections are reported.}
\end{figure}

Several stars from our catalogue have several abundances because these targets were originally observed 
by two spectrographs (we selected the best spectrum between both as explained before). This allows us to test 
the internal accuracy of our automatic method. 
For that purpose, we selected stars observed both with UVES and HARPS, resulting in a subsample of 
117 dwarfs and giants (only very few stars were observed with HARPS and FEROS or UVES and FEROS, 
and we rejected these repeats because of insufficient statistics). This subsample covers a wide range of parameters, i.e. 
$4500<\tef<6700\,$K, $2.5<\logg<4.9\,$\cms, and $-0.9<\metah<+0.4\,$dex. We compared both lithium abundances 
and upper limits (see \figurename{~\ref{compare_harps_uves}}). 

On the one hand, we see that the 72 lithium measurements are in a very good agreement between both spectrographs 
with a mean difference of $0.05\,$dex and a standard deviation equal to $0.18\,$dex. This dispersion can 
be due to the standard deviation of the difference of the AMBRE effective temperatures between 
both spectrographs ($107\,$K). The small bias can be related to the biases in $\tef$ and $\logg$, $17\,$K 
and $-0.11\,$\cms, respectively between both spectrographs. The bias in $\metah$ is null. On the other hand, we see that 
the 43 upper limits of HARPS are systematically higher than the UVES upper limits. This can be well explained by the 
fact that, for this subsample, the average $\snr$ of HARPS targets is three times lower than for UVES; 
the detection limits are partly governed by the $\snr$ values. We also have two stars 
with lithium detection in UVES and upper limit with HARPS and a good consistency is found between them. 
We then conclude that our automatic method is robust since it is able to recover with good 
precision the lithium abundances when a star is observed by two different spectrographs. 

\subsection{Comparison with independent Li abundances}

\subsubsection{\citet{ramirez_2012}}

We first compare the AMBRE LTE lithium values with those of \citet{ramirez_2012} derived 
from UVES and HARPS spectra. The AMBRE/Li catalogue contains 74 dwarf stars with spectra already 
analysed by the authors, \emph{i.e.} $11\%$ of their sample. 
The atmospheric parameters of these stars in common are within $5300<\tef<6300\,$K, 
$3.6<\logg<4.7\,$\cms~and $-1.0<\metah<+0.2\,$dex. \citet{ramirez_2012} also adopted a different 
spectral synthesis code (MOOG2010; \citealt{sneden_1973}) together with a different line list
to determine their lithium abundances. 
The comparison of the derived lithium abundances is shown in 
\figurename{~\ref{compare_delgado}} (left panel) for 59 stars, excluding upper limits. 
We see a good agreement between both studies with small bias and dispersion 
of $-0.04$ and $0.16\,$dex, respectively. This dispersion can be well 
explained by the dispersions of the difference in the adopted $\tef$, $\logg$ and $\metah$ 
between both groups that we estimated to be $95\,$K, $0.18\,$\cms~and $0.08\,$dex, respectively. 
We also checked the measured upper limits of 25 stars
and a very good consistency is also found.

\subsubsection{\citet{delgado_mena_2015}}

We compare the AMBRE LTE lithium values with those derived from HARPS 
spectra published by \citet{delgado_mena_2015}. The AMBRE/Li catalogue contains 194 stars with HARPS 
spectra already analysed by \citet{delgado_mena_2015}, \emph{i.e.} $55\%$ of their sample. 
Apart from their line list, their adopted procedure to derive the lithium abundances is very 
close to that of \citet{ramirez_2012}. The atmospheric parameters of these stars in common are within $5600<\tef<5900\,$K, 
$3.6<\logg<4.7\,$\cms~and $-1.0<\metah<+0.4\,$dex. 
The comparison of the derived lithium abundances is shown in 
\figurename{~\ref{compare_delgado}} (middle panel) for 124 stars, excluding upper limits. 
We see a very good agreement between both studies with no bias and a small dispersion 
of $0.10\,$dex. This dispersion can be well explained by the dispersions of the 
difference in the adopted $\tef$, $\logg$ and $\metah$ between both groups that we estimated to be 
$62\,$K, $0.17\,$\cms~and $0.06\,$dex, respectively. 
We clearly see, however, that this dispersion increases for lower $\ali$ values. At 
lower abundance and fixed other parameters, the blends contribution to the line profile starts 
to be stronger (particularly for cool stars) and a different blend treatment between both methods 
could explain such higher dispersion. We also checked the measured upper limits of 70 stars 
and a very good consistency is found again.

\subsubsection{\citet{delgado_2015}}

\begin{figure*}
\centering
\includegraphics[width=0.32\linewidth]{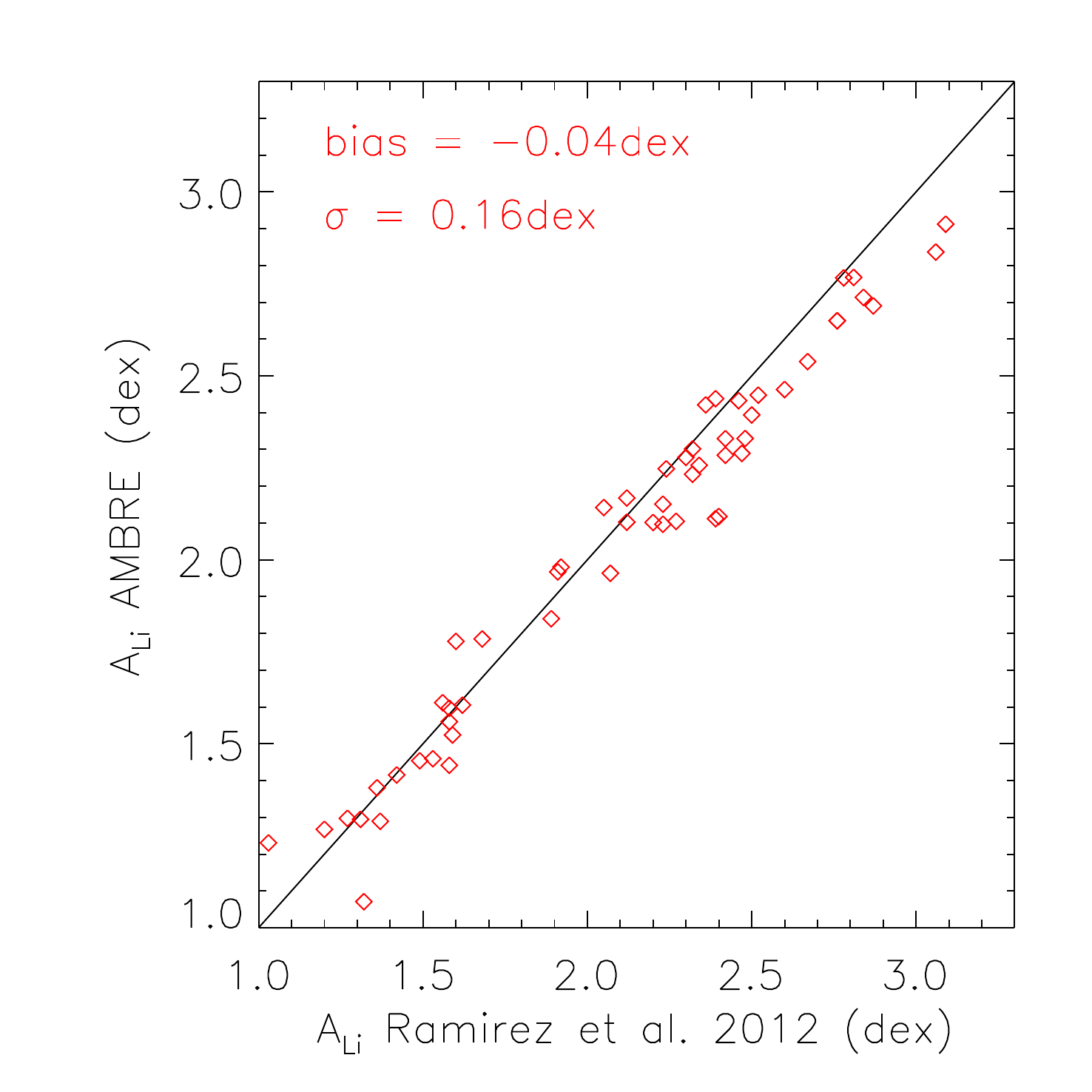}
\includegraphics[width=0.32\linewidth]{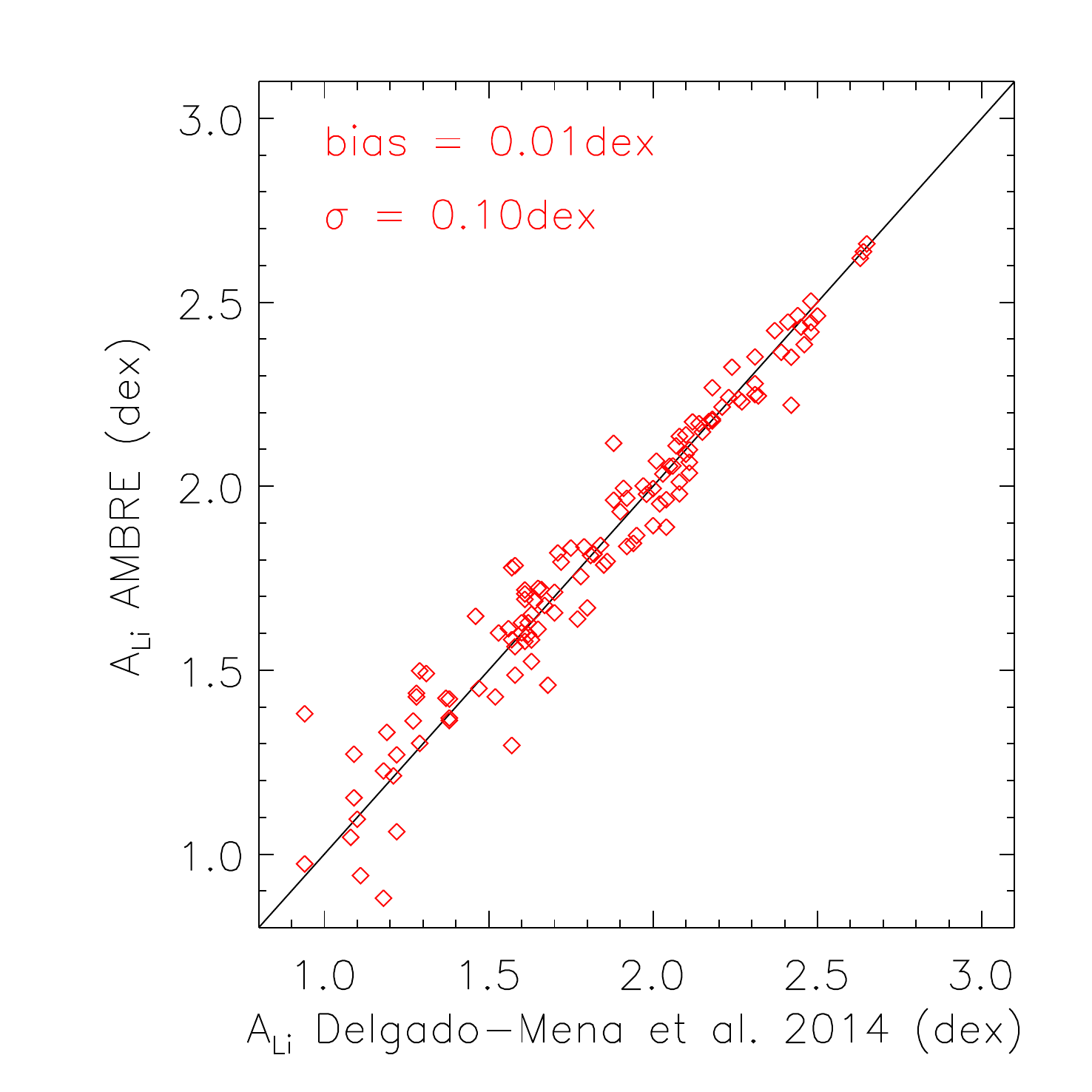}
\includegraphics[width=0.32\linewidth]{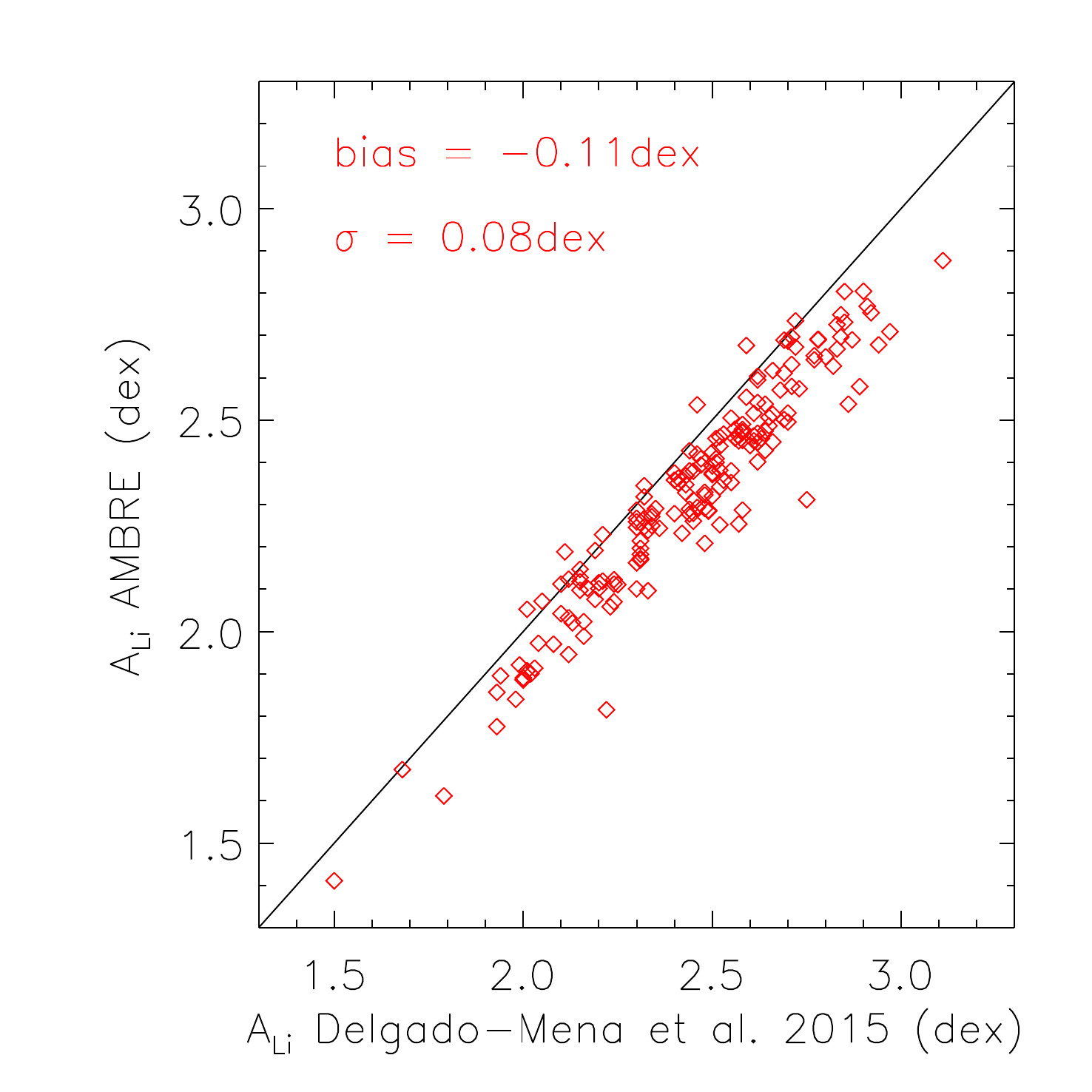}
\caption{\label{compare_delgado} Comparison between the LTE lithium abundances 
of the AMBRE/Li catalogue and the studies of \citet{ramirez_2012}, 
\citet{delgado_mena_2015}, and \citet{delgado_2015}.}
\end{figure*}

We finally compare the AMBRE LTE lithium values with those of \citet{delgado_2015} 
derived from HARPS spectra. There are 200 hot stars, \emph{i.e.} $61\%$ of their sample. 
The atmospheric parameters of these stars in common are within $5900<\tef<6700\,$K, 
$3.5<\logg<4.5\,$\cms~, and $-1.0<\metah<+0.4\,$dex. The comparison of the derived lithium 
abundances is shown in \figurename{~\ref{compare_delgado}} (right panel) for 179 stars, 
excluding upper limits. We see a good agreement between both studies with a bias of $-0.11\,$dex 
and a small dispersion of $0.08\,$dex. This dispersion can be well explained by the dispersions of the 
difference in the adopted $\tef$, $\logg$ and $\metah$ between both groups that we estimated to be, 
$78\,$K, $0.19\,$\cms~and $0.07\,$dex, respectively. The bias in $\tef$ is not constant 
and increases with $\tef$. We also checked the measured upper limits of 21 stars and a very good 
consistency is found.

\subsection{Lithium abundance of the Gaia-benchmark stars}\label{gaia_bench_abund}

\begin{figure*}
\centering
\includegraphics[width=1.0\linewidth]{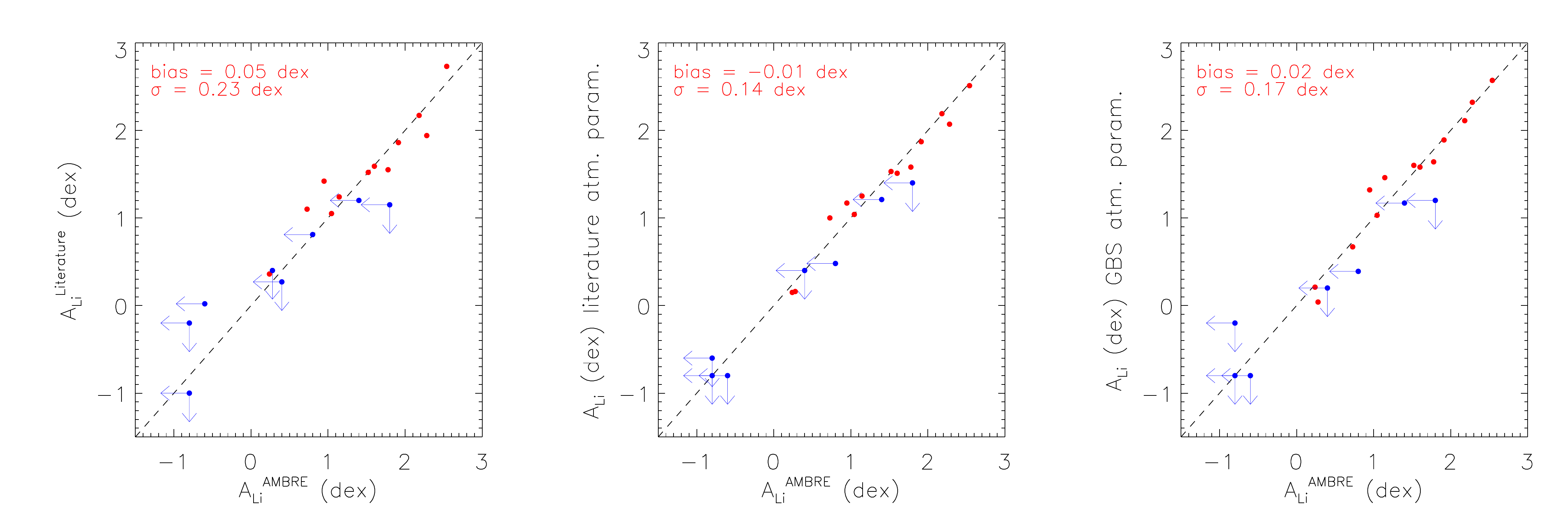}
\caption{\label{compare_bench} LTE lithium abundances for a set 
of 20 Gaia-benchmark stars. \emph{Left panel:} literature abundances compared to the 
AMBRE/Li determinations. \emph{Middle panel:} lithium abundances derived from the AMBRE 
spectra and adopting the literature atmospheric parameters compared to the AMBRE/Li determinations. 
\emph{Right panel:} lithium abundances derived from the AMBRE 
spectra and adopting the Gaia benchmark stars (GBS) atmospheric parameters compared to 
the AMBRE/Li determinations. Lithium abundances 
are shown in red, while upper limits are represented by blue arrows. Mean differences and standard 
deviations for the lithium abundance are reported. Stars and references of the literature values are 
the following: $\boldsymbol\epsilon$ \textbf{Eri}, $\boldsymbol\beta$ \textbf{Vir} \citep{gonzalez_2010}; 
$\boldsymbol\alpha$ \textbf{Cen A}, \textbf{18 Sco} \citep{baumann_2010}; \textbf{HD22879} \citep{nissen_2012}; 
\textbf{Sun} \citep{grevesse_2007}; $\boldsymbol\tau$ \textbf{Cet} \citep{takeda_2005}; $\boldsymbol\alpha$ 
\textbf{Cen B} \citep{chmielewski_1992}; $\boldsymbol\mu$ \textbf{Ara}, $\boldsymbol\delta\,$ \textbf{Eri}, 
$\boldsymbol\beta$ \textbf{Hyi} \citep{bruntt_2010}; \textbf{Procyon}, \textbf{HD49933} \citep{ramirez_2012}; 
\textbf{HD84937} \citep{lind_2013}; \textbf{HD140283} \citep{asplund_2006}; $\boldsymbol\epsilon$ \textbf{For} 
\citep{lebre_1999}; \textbf{Arcturus} \citep{reddy_2005}; $\boldsymbol\epsilon$ \textbf{Vir}, $\boldsymbol\alpha$ 
\textbf{Tau} \citep{mallik_1999}; and \textbf{HD107328} \citep{brown_1989}.}
\end{figure*}

As an additional check of our automatic procedure, we identified several Gaia-benchmark stars \citep{jofre_2013} in the AMBRE/Li
catalogue. This identification 
was performed with the coordinates and TARGNAME identifier, resulting in 
20 stars. We also measured $\ali$ for these same identified stars 
from the AMBRE spectra, but adopted the atmospheric parameters presented by recent studies 
reporting lithium abundances. We emphasize the fact that the literature studies 
are characterized by different spectral resolutions, $\snr$, atomic and molecular 
treatment, and also different techniques 
(e.g. spectral fitting and equivalent width analysis) for the abundance derivation. All these facts 
can be responsible for most of the (small) differences observed with respect to the literature. 
As presented in \figurename~\ref{compare_bench}, we clearly see in the left panel that the comparison with published 
LTE lithium abundance is good. We measure a small bias (defined as the mean difference) of $0.05\,$dex and a 
dispersion $\sigma=0.23\,$dex that can easily be explained by the different atmospheric parameters 
(and particularly $\tef$) adopted in AMBRE and these literature studies. Indeed, when comparing 
the AMBRE/Li abundances with the lithium abundances derived 
using the atmospheric parameters from the literature (middle panel), the agreement becomes very satisfactory 
with no bias and a smaller dispersion $\sigma=0.14\,$dex. Finally, when adopting the effective temperature and 
surface gravity of \citet{heiter_2015gaia} and metallicity of \cite{jofre_2014_metal}, 
we see in \figurename~\ref{compare_bench} (right panel) that the derived 
lithium values are once again consistent with the AMBRE/Li 
abundances. The atmospheric parameters that are derived by \citet{heiter_2015gaia} 
and \cite{jofre_2014_metal} are very similar to the independent determinations by the AMBRE project. 
These comparisons again confirm that the lithium 
abundances of the AMBRE/Li catalogue are reliable for a scientific exploitation.

\section{Tracing the lithium evolution in the Milky Way}\label{discussion}

In this section, we propose to study the evolution of lithium enrichment 
in the Milky Way using the AMBRE catalogue of lithium abundances. We first 
investigate the stellar lithium content as a function of the metallicity used 
as a time tracer to constrain the abundance of the ISM in which 
they were formed. Then, we study the lithium enrichment in both the galactic 
thin and thick discs. To do this, we built a \emph{working} sample from the 
catalogue presented in Sect~\ref{catalouge}, rejecting stars with upper and 
lower limits in their lithium abundances. As we also want to avoid any possible 
lithium abundance variations caused by stellar evolution during the late stages, 
we only consider dwarf stars (defined hereafter as $\logg\ge3.7\,$\cms). 
Finally, as the catalogue covers a wide range in effective temperature 
and metallicity, we selected targets with available NLTE corrections to 
compare the right lithium  abundance correctly in a rather different type of stars. 
As a result, our \emph{working} sample is composed of $3\,077$ stars with 
minimal and median $\snr$ value of 15 and 82, respectively.

\subsection{Lithium evolution with the metallicity}\label{sect_evol_li_mh}

The classical method to study the evolution of a chemical element in the Milky Way is 
to study its behaviour with the metallicity, used as a proxy of age (at least 
for $\metah<+0.0\,$dex). The main problem with the study of the Galactic lithium evolution with 
metallicity is that lithium is depleted in stellar interiors along the life 
of the star, and some Li production can even occur at specific evolved stages and cannot be 
assumed to be representative of the Li abundance of the ISM material from 
which the star was formed. This phenomenon leads to a broad spread in $\ali$ whatever the metallicity.
The metallicity in our catalogue covers a broad range, from $\metah=-3\,$dex 
for halo stars to $\metah=+0.5\,$dex for the richest disc stars. The goal of the present 
section is to investigate the ISM lithium abundance variation along this wide metallicity range. 
To accomplish this, we selected hot stars from the \emph{working} sample with $\tef>5\,600\,$K to reject the coolest dwarf stars with deeper convective zone (higher lithium destruction), 
thereby creating  a \emph{clean} sample of $2\,310$ stars.

\subsubsection{The AMBRE stars of the Spite plateau}

For more than three decades hot metal-poor stars have been known  to exhibit a rather constant 
lithium abundance with metallicity ($\ali\sim2.2\,$dex; \citealt{spite_1982}). This 
plateau was first interpreted as the primordial lithium abundance, but since the recent 
results of the \emph{Wilkinson Microwave Anisotropy Probe} (WMAP) mission, 
the lithium abundance of the Standard Big Bang Nucleosynthesis (SBBN) has been revised to $\ali\sim2.6\,$dex 
\citep{spergel_2003}. This discrepancy is know as the \emph{lithium problem}, 
and several studies proposed different depletion mechanisms to explain such a difference, 
(\emph{e.g.} \citealt{pinsonneault_1999}).

\begin{figure}
\centering
\includegraphics[width=1.0\linewidth]{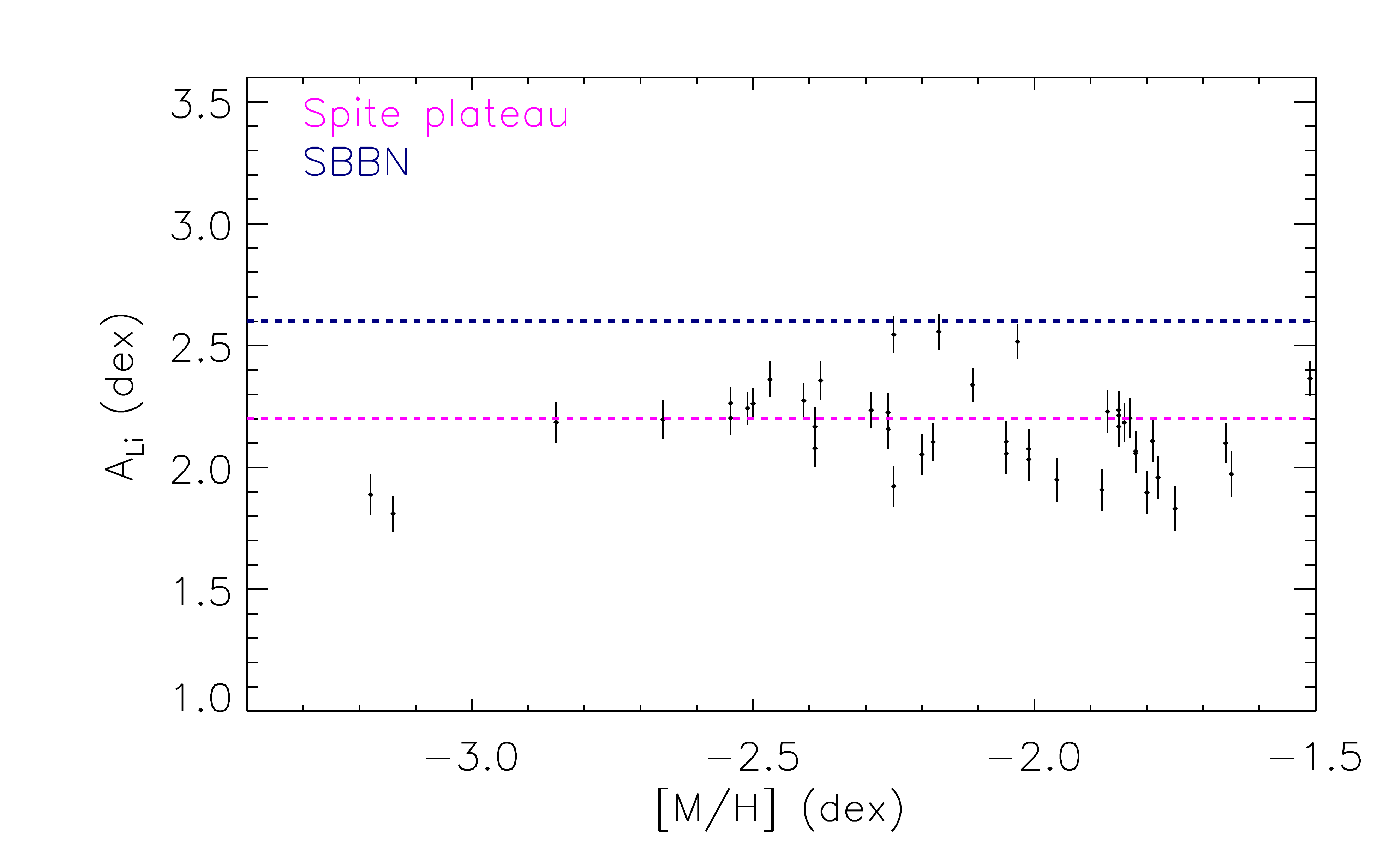}
\caption{\label{spite_plateau} $\ali^{\text{NLTE}}$ of the AMBRE/Li catalogue stars 
as a function of $\metah$ for $\metah<-1.5\,$dex. We overplotted the typical 
SBBN and Spite plateau lithium abundances in blue and magenta, respectively.}
\end{figure}

We present the NLTE lithium abundances of the 44 metal-poor stars 
($\metah<-1.5\,$dex) of the \emph{clean} sample  in \figurename~\ref{spite_plateau}. First, we clearly see that stars with 
$-3.0<\metah<-1.5\,$dex show a rather constant lithium abundance ($\langle\ali\rangle=2.08\,$dex) with 
a typical dispersion of $0.22\,$dex. We thus confirm that the mean lithium abundance of the most 
metal-poor stars in the Galaxy is lower by $\sim0.4/0.5\,$dex with respect to the SBBN value. 
Second,   we found two stars with $\metah<-3\,$dex 
and typical lithium abundances that are much lower than the Spite plateau ($\ali^{\text{AMBRE}}=1.81, 1.88\,$dex). 
These stars have already been analysed by \citet{sbordone_2010} with the same UVES data and our lithium abundances are 
consistent with these within $2\sigma$ errors ($\ali^{\text{Sbordone}}=2.10, 1.65\,$dex). The differences between 
the spectroscopic AMBRE $\tef$ and the photometric temperatures of \citet{sbordone_2010} 
are $-413\,$K and $+528\,$K, respectively, and can explain such differences in abundances. 
A reason why these stars show lower lithium abundances 
compared to the Spite plateau could be that they suffered from a subsequent lithium depletion, as proposed 
by \citet{sbordone_2010}. Also, we find three stars with lithium abundances that are higher 
that $\ali>2.5\,$dex ($\ali^{\text{AMBRE}}=2.54, 2.51, 2.56\,$dex). These targets were already 
observed by \citet{lind_2009_ngc} and our lithium abundances are consistent with these targets within $2\sigma$ errors 
($\ali^{\text{Lind}} 2.31, 2.19, 2.16\,$dex). The differences between the AMBRE $\tef$ and the photometric $\tef$ of 
\citet{lind_2009_ngc} are $+567\,$K, $+361\,$K, and $+407\,$K, respectively, and can again explain such 
differences in abundances. \citet{lind_2009_ngc} used GIRAFFE/HR15 data 
($R\sim19\,200$) contrary to UVES data in our study.

\subsubsection{ISM lithium abundance \emph{v.s.} $\metah$}

\begin{figure*}
\centering
\includegraphics[width=0.8\linewidth]{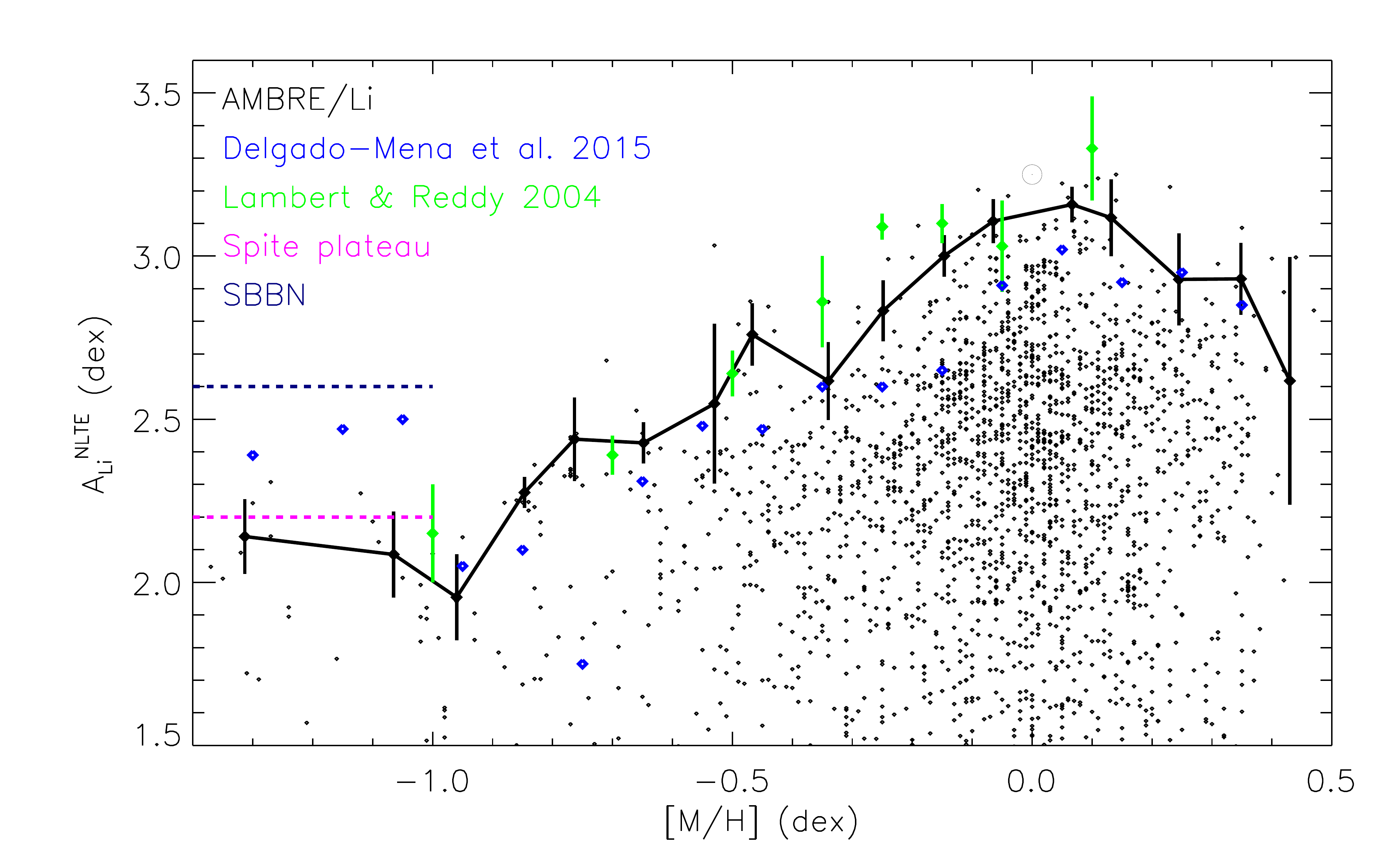}
\caption{\label{li_feh} Mean lithium value for the 6 stars with the highest 
lithium abundance in each metallicity bin as a function of the metallicity. The AMBRE relation is shown in black, 
while the relation of \citet{lambert_2004} and \citet{delgado_2015} are 
overplotted in green and blue, respectively. We overplotted in black all the AMBRE 
individual lithium measurements used to build the black curve together with the solar meteoritic 
lithium abundance.}
\end{figure*}

As emphasized previously, lithium depletion mechanisms seem to occur at 
all metallicities, especially for $\metah>-1.5\,$dex. To study the lithium 
abundance in the ISM, one has  to consider the upper 
envelope of the lithium distribution that reflects the initial condition of the interstellar medium to a greater extent, \emph{i.e.} stars with no lithium depletion. 
For that purpose, we followed the same approach as \citet{lambert_2004} and \citet{delgado_2015}. 
 We regularly binned the data 
of the \emph{clean} sample with a step $\Delta\metah=+0.10\,$dex for the range $-1\le\metah\le+0.5\,$dex to trace $\ali$ as a function of the metallicity. 
We also considered two bins with $\Delta\metah=+0.20\,$dex on the domain $-1.4<\metah<-1.0\,$dex. This 
final subsample consists of a total of $2\,264$ stars. Errors on lithium abundances are 
typically $0.1\,$dex. For all the metallicity bins, we then chose the six stars with the highest 
NLTE abundance in lithium and computed their mean $\ali$. We adopted this amount of stars per bin as 
in previous studies, but we confirm that our results are robust when selecting 4, 6, or 8 stars. For a 
given bin, the error bar is finally given by the standard deviation of the 6 $\ali$ measurements. 
The AMBRE relation shown in \figurename{~\ref{li_feh}} is stable with both the metallicity binning and the cut in $\tef$ 
and $\logg$. We also emphasize the fact that it is the first time that this type of study is carried out with 
such a large statistical sample size and a perfectly homogeneous catalogue over a very wide metallicity range.

It can be seen in \figurename{~\ref{li_feh}} that for $\metah<-1\,$dex, the AMBRE relation 
is consistent with the Spite plateau, as with the point of \citet{lambert_2004}. However, the three 
points from \citet{delgado_2015} show a strong disagreement with our data, probably 
because their last two bins are composed of only one star (see their Fig. 5). 
Moreover, we clearly measure an increase of the maximum lithium abundance in the range 
$-1.0\lesssim\metah\lesssim+0.1\,$dex, reaching $\ali=3.16\,$dex in the domain $+0.0\le\metah\le+0.1\,$dex, which is close to 
the meteoritic abundance value. It can be seen that our AMBRE curve is in a rather good 
agreement with \citet{lambert_2004}, while the relation of \citet{delgado_2015} is systematically 
lower, possibly due to sample selection effects. At super-solar metallicity, 
the maximum lithium abundance decreases clearly by a factor $0.55\,$dex in the range $+0.00<\metah<+0.50\,$dex. 
We are confident with this behaviour since it is constrained by five data points. 

For $\metah>+0.0\,$dex, we visually checked that the quality of the fit between the observed and synthetic spectrum is good 
for the six stars with the highest lithium abundance in each metallicity bin. The average $\tef$ of these six 
richest stars tends to increase with the metallicity, so that in this situation internal destruction of lithium 
does not seem to be the reason for the decrease. In addition, the errors in $\tef$ and $\ali$ tend 
to decrease with the metallicity, while for $\metah$ the average error is very constant.

Several Galactic chemical evolution models (GCE) tried to reproduce the rise of lithium with metallicity 
by considering the contributions of several sources: spallation by Galactic cosmic rays (GCC), the $\nu-$process by the 
core-collapse supernovae (CCSN), novae, low-mass giants and asymptotic giant branch stars (AGB), 
starting from a primordial lithium value (SBBN, $\ali=2.6\,$dex). \citet{fields_1999} developed a  GCE, considering only the 
GCC and CCSN contributions; the model cannot fit the meteoritic lithium abundance because the author did not 
include stellar sources. Later, \citet{romano_2001} 
proposed a model with the five previously quoted sources and a dominant low-mass star contribution 
that fits the meteoritic value rather well. The same year \citet{travaglio_2001} concluded that the major contribution 
comes from AGBs, while the role of low-mass giants, novae, and CCSNs is weak. Carbon-rich AGB stars were proposed by 
\citet{alibes_2002} as the main source of lithium at $\metah>-0.5\,$dex. 

More recently, \citet{prantzos_2012} 
provided a new GCE model for the light elements Li, Be, and B, considering all possible sources of lithium: 
SBBN, AGBs, low-mass giants, novae, GCRs, and CCSN. He found that i)
the two best-known sources of Li, namely SBBN and GCR, can provide about 10\% and 20\%of the solar Li, respectively, 
leaving the remaining 70\% for a stellar source, and ii) current yields from all the aforementioned stellar sources
fail to provide the remaining amount  by  factors $\sim$5-10, thus calling for a considerable reassessment of Li production in stars.
He also argued that the uncertainties in stellar Li yields and the amount of Li depletion in stellar envelopes make it difficult to make
any meaningful comparison between GCE models and Li observations.

The behaviour of $\ali$ at supersolar metallicities reported here, namely its decrease  for $\metah>+0.0\,$dex, 
is  not reproduced by the aforementioned models with
the possible exception of \citet{fields_1999}. However, that model neglected the main stellar sources of Li, thus  failing to reproduce
the solar Li value, and the reasons for the obtained  decrease is not clear. The Li decrease revealed by the AMBRE/Li data analysis
is unique (no other element, besides D, displays a decreasing abundance) and requires further investigation on theoretical grounds 
and an independent observational confirmation.

\subsection{Lithium in the thin and thick discs}

It is now well established that the Milky Way disc has two major components. Indeed, 
the presence of a thin and thick disc has been revealed by many methods, for example by stellar counts 
\citep{gilmore_1983}, dynamically \citep{bensby_2014} and 
chemically by $\alffe$ ratio abundances (\emph{e.g.} \citealt{Adibekyan2011, ges_disc_recio_2014}), suggesting 
at least two different evolutionary paths in the history of the Milky Way disc formation. 

In this context, 
two recent works attempted to study both discs in order to detect a possible distinct chemical 
lithium evolution in the thin and thick discs. \citet{ramirez_2012} 
observed that the thin disc stars show a rather high lithium content, marked by an enrichment in $\ali$ with 
increasing $\metah$, while the maximum thick disc lithium abundances are constant with 
$\metah$ and very close to the Spite plateau. By decomposing the thin disc in age slices
\footnote{\citet{ramirez_2012} determined the ages thanks to an isochrone fitting method and distances 
from Hipparcos parallaxes.}, \citet{ramirez_2012} 
also showed that the lithium abundance envelope of the thin disc is in continuity with that observed in the thick disc. 
These authors applied a kinematical criteria to separate thin and thick stars. In a more 
recent work, \citet{delgado_2015} proposed that the 
thick disc lithium abundances decrease with increasing metallicity in contrast with \citet{ramirez_2012}, 
while the thin disc shows higher lithium content than the thick disc (as already revealed by \citet{ramirez_2012}). 
 \citet{delgado_2015} disentangled both discs thanks to the kinematical criteria of \citet{ramirez_2012}, 
as well as by searching for gaps in the $\alffe$ distribution for a given metallicity range. \citet{delgado_2015} 
concluded that these two thin to thick disc separations do not influence their results.

We used the AMBRE/Li catalogue to search for a possible distinction in lithium between both 
Galactic discs. The high statistics and homogeneity of this catalogue offer a good opportunity to perform 
such an analysis. In contrast to the two previous quoted studies, we are able to provide a new view on this 
problem. In order to chemically separate thin to thick disc stars, we built a \emph{clean} sample from the 
\emph{working} sample presented at the beginning of this section, composed of stars with the best atmospheric 
parameters, \emph{i.e.} $\text{QUALITY}\_\text{FLAG}=0$ and $\snr>150$ to establish a robust chemical separation. 
We also rejected stars with $\metah\le-1.25\,$dex that are probably halo members and weakly representative of the thin and 
hick discs in terms of metallicity. The resulting \emph{clean} sample is composed of $363$ very well parametrized stars. 
It is used to define our methodology for the thin/thick disc classification.

\begin{figure}
\centering
\includegraphics[width=1.0\linewidth]{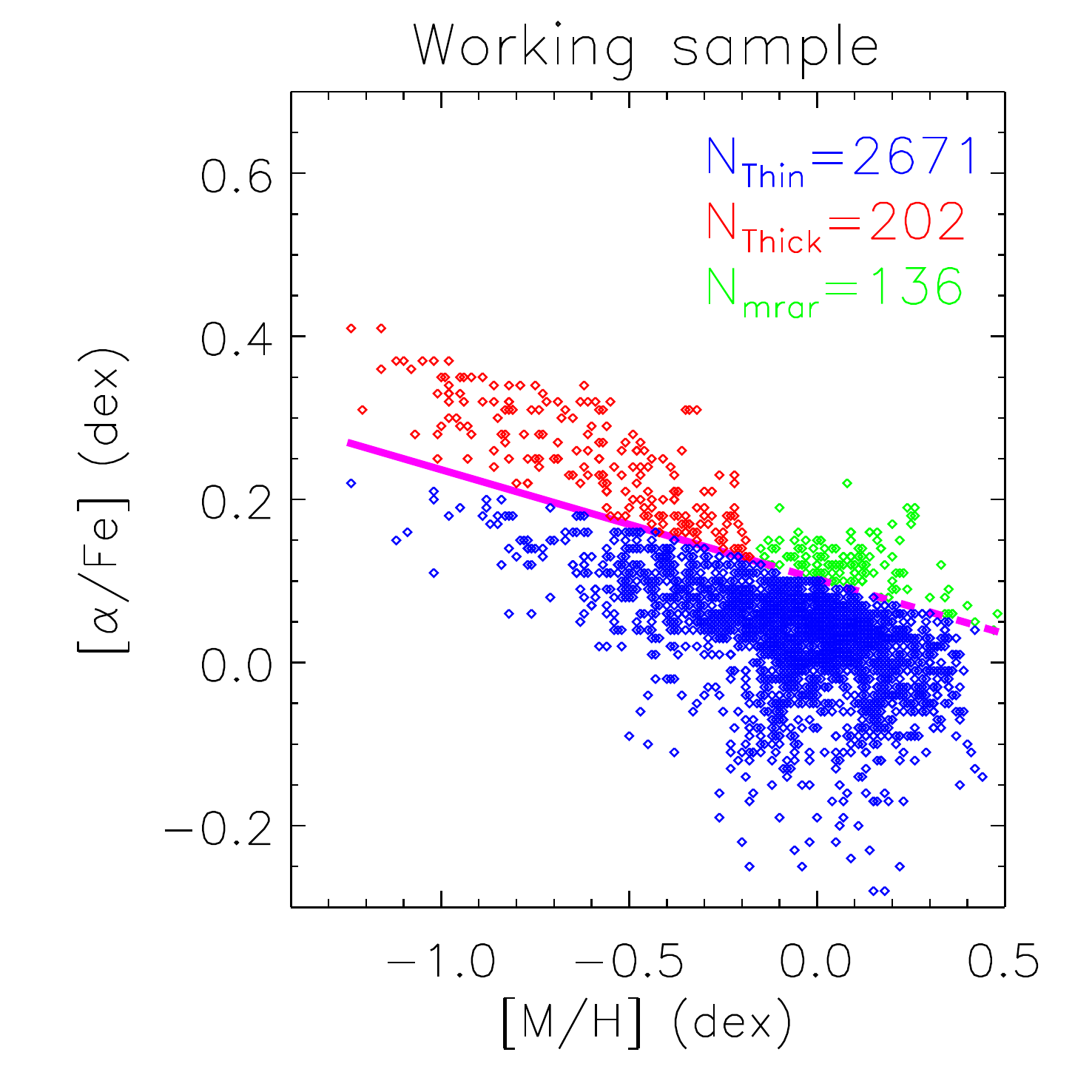}
\caption{\label{thin_to_thick} $\alffe$ ratio as a function of $\metah$ of the AMBRE/Li 
stars for the \emph{working} sample ($3\,009$ stars). The full magenta line shows the 
thin to thick disc separation. The magenta dashed line shows the extrapolated separation 
for $\metah>-0.15\,$dex. The thin disc stars are colour-coded in blue, while thick disc 
members are in red. The metal-rich $\alpha$-rich stars are shown in green.}
\end{figure}

We used the $\alffe$ ratio and $\metah$ provided by the AMBRE project to disentangle the two discs.
We followed the same procedure as in \citet{guiglion_2015}. We decomposed the $\metah$ distribution of our \emph{clean} sample 
in metallicity bins, searching visually for gaps in the corresponding $\alffe$ distribution. We were able to separate 
two components between $-1.1\le\metah\le-0.15\,$dex. We linearly fitted 
these gaps to separate high-$\alffe$ stars (thick disc with $-1.25\le\metah\le-0.15\,$dex) from low-$\alffe$ stars 
(thin disc with $-1.25\le\metah\le-0.50\,$dex), following an extrapolated separation on the domain $-1.25\le\metah\le+0.50\,$, 
as shown in \figurename~\ref{thin_to_thick}. Metal-rich $\alpha$-rich stars with $\metah>-0.15\,$dex and $\alffe$ above 
the separation were rejected because they are too metal-rich compared to the classical definition of the thick disc. 
We finally applied this separation to the whole \emph{working} sample on the same metallicity 
domain ($3\,009$ stars, $89\%$ thin disc, $7\%$ thick disc, and $4\%$ metal-rich $\alpha$-rich stars). The thick disc stars are in the minority compared to the thin disc stars; this is probably because the AMBRE 
content of the ESO archive is not optimized to target thick disc stars as for 
example in the Gaia-ESO Survey.

\begin{figure*}
\centering
    \centerline{\includegraphics[width=21cm]{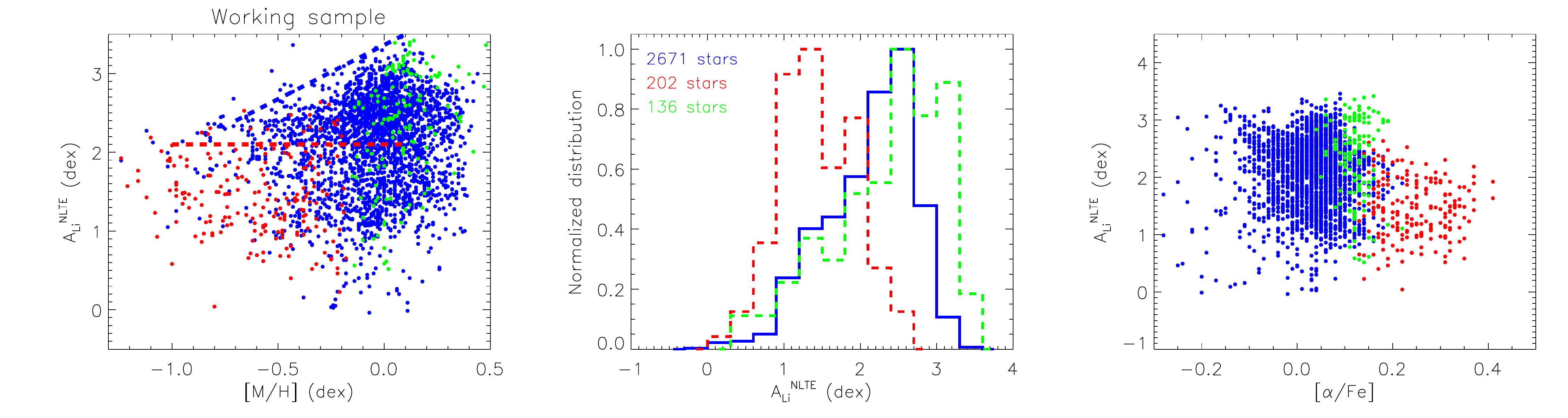}}
\caption{\label{thin_to_thick_res} $\ali^{\text{NLTE}}$ as a function of $\metah$ (\emph{left}) and 
its normalized distribution (\emph{center}) of the AMBRE/LI stars for the \emph{working} sample. 
The thin disc stars are plotted in blue, thick disc members in red and metal-rich $\alpha$-rich 
stars in green. The upper envelope of both discs from \citet{ramirez_2012} 
are shown as dashed lines (bottom, left). We also show $\ali^{\text{NLTE}}\,vs.\,\alffe$ in the 
\emph{right} panel (Spearman’s rank correlation coefficient equal to -0.03 and -0.04 
for the thin and the thick disc, respectively).}
\end{figure*}

Based on this thin to thick disc characterization, the lithium content in each disc is 
shown in \figurename~\ref{thin_to_thick_res}, for the \emph{working} sample.

\begin{enumerate}

\item [-] The AMBRE thick disc stars exhibit lower abundances of lithium than those in the thin disc 
and the highest lithium abundances seem to increase slightly with $\metah$, around $\ali\sim2-2.2\,$dex. In addition, 
the normalized distribution of $\ali$ in the thick disc clearly shows two peaks at 
$\sim2.2$ and $\sim1.2\,$dex, respectively. The higher value seems to correspond to an extension of the Spite plateau, 
while the lower value is certainly due to internal lithium destruction. Our results for the thick disc 
are slightly different from the study of \citet{ramirez_2012} described earlier and contradict the 
decrease of lithium with $\metah$ in the thick disc shown by \cite{delgado_2015}. 

\item [-] As shown in \figurename~\ref{thin_to_thick_res}, the lithium abundance 
for the thin disc stars increases with metallicity, reaching the highest 
values around solar metallicity and decreases at super-solar metallicity (as already noted in 
Sect~\ref{sect_evol_li_mh}). The normalized $\ali$ distribution of the thin disc is 
characterized by a single peak ($\ali\sim2.5\,$dex) with a tail towards 
lower $\ali$. The rise of the upper lithium envelope of the thin disc is in agreement 
with \citet{ramirez_2012} (as shown in the \figurename~\ref{thin_to_thick_res}), while 
we confirm the decrease of $\ali$ at super-solar metallicity, suggested in data from \citet{delgado_2015}.

\end{enumerate}

We also observe that the lithium abundance distribution of the metal-rich $\alpha$-rich stars 
has a peak at $\ali\sim3\,$dex, with a tail extending to lower lithium abundances. The higher peak 
compared to the thin and thick disc could indicate a different ISM enrichment history. In addition, 
a link with the Galactic bulge is not obvious because of numerous stars with $\ali>2.5\,$dex, compared 
to bulge studies pointing out stars with $\ali<2.5\,$dex \citep{pompeia_2002, barbuy_2010}. 

Our study is based on a thin to thick disc separation using a chemical criterion in contrast to \citet{ramirez_2012} who applied a dynamical selection. Moreover, as suggested 
by \citet{ramirez_2012}, we emphasize the fact that these results can suffer from biases in 
ages, masses, and metallicity between both discs, and then show different degrees of lithium 
depletion or atomic diffusion, instead of real lithium enrichment. However, the agreement is real between 
both studies.

In \figurename~\ref{thin_to_thick_res} (\emph{right panel}), it can be suggested that 
in the thick disc, the upper envelope of the stellar lithium content is correlated 
with the $\alffe$. As the $\alffe$ ratio is well correlated with the age in the thick disc 
(see for example \citet{Haywood_2013}), we directly see the evolution of lithium with time. 
On the other hand, no clear trend of $\ali$ is observed in the thin disc with $\alffe$.

From the Galactic chemical evolution point of view, our finding confirms that the ISM where the 
thick disc stars were formed has not been significantly enriched from the Spite plateau abundance value 
($\ali\sim2.2\,$dex). In the  model of \cite{prantzos_2012} (see e.g.  Fig. 16 of that work), the ISM has been weakly enriched by CCSNs and GCRs at 
low metallicity ($\metah<-0.6\,$dex), whereas low-mass stars are able to enrich sensitively the ISM only at higher 
metallicity ($\metah.6\,$dex), corresponding to the thin disc phase. However, the 
observed double-branch behaviour of Li cannot be interpreted in the framework of the simple 1-zone model of \cite{prantzos_2012} or other previous GCE models.
For the moment, we can only infer that the low-mass stellar sources of Li  have only played a key role in the ISM enrichment of the thin disc. 

\begin{figure}
\centering
\includegraphics[width=1.0\linewidth]{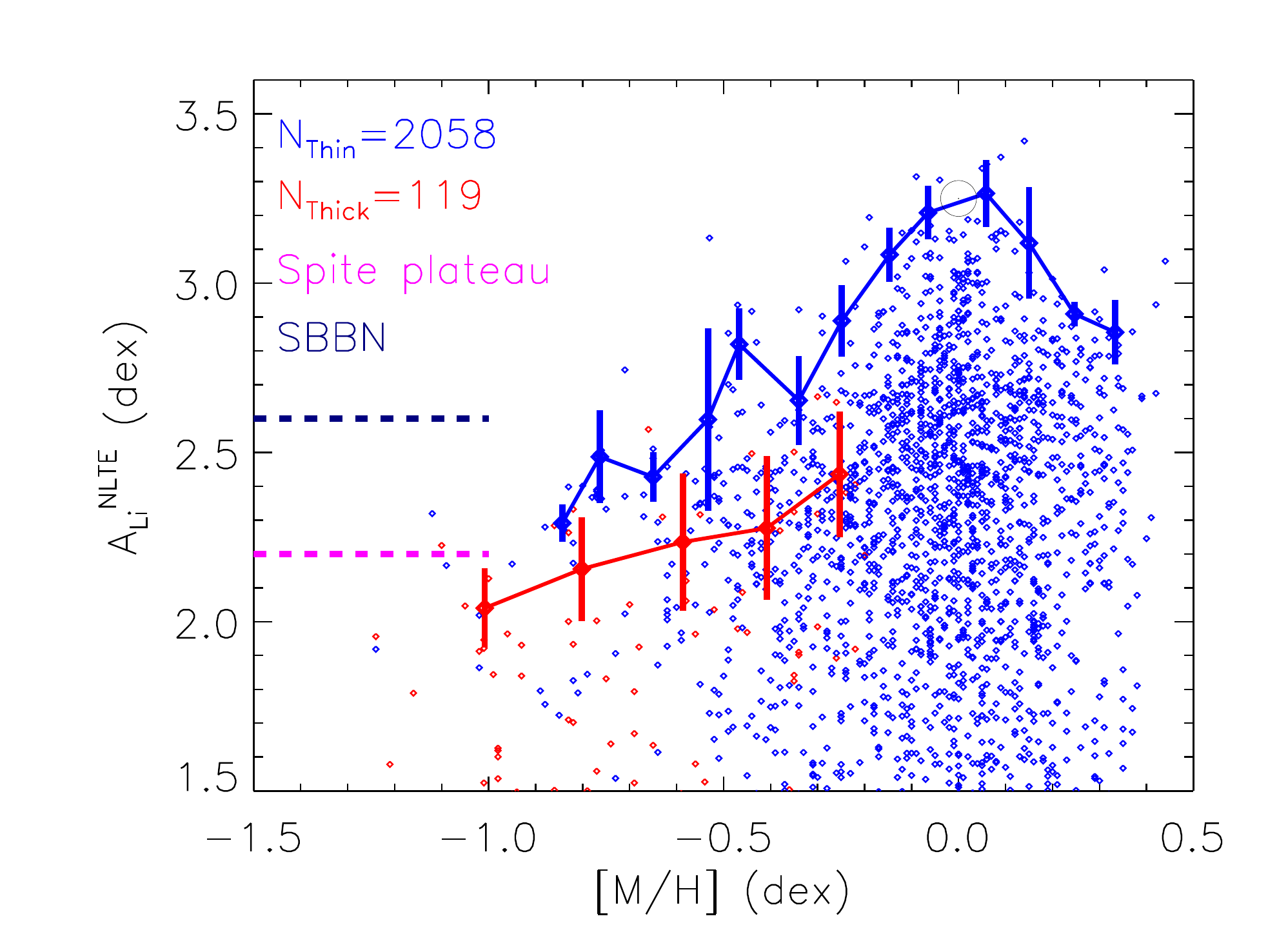}
\caption{\label{thin_to_thick_res_max_ab} Same figure as \figurename~\ref{li_feh}, 
but taking into account the thin (in blue) to thick (in red) characterization.}
\end{figure}

Finally, we repeated the analysis presented in Sec~\ref{sect_evol_li_mh} to 
investigate the lithium content of the ISM in which the thin and thick disc were formed. 
We removed the stars with $\tef<5600\,$K as in Sect~\ref{sect_evol_li_mh}, leading to fewer stars compared to 
\figurename~\ref{thin_to_thick_res}. We show our results in \figurename~\ref{thin_to_thick_res_max_ab}. 
While the thick disc shows a slightly increasing maximum abundance in spite of the cut in $\tef$, 
the thin disc shows a typical increase of lithium abundance with $\metah$ to lithium meteoritic abundance 
at solar metallicity. The important highlight is that we are able to show the lithium decrease 
at super-solar metallicities in the thin disc. The two distinct relations in the thin and thick disc indicate then that both discs are characterized by two distinct lithium enrichments.

Recent chemical evolution models manage to produce a "thick disc" (older than $\sim$8-9\,$Gyr$) 
through radial migration {of stars from the inner disc. This leads naturally to two branches in the 
$[\text{O}/\text{Fe}]-\feh$ space corresponding to the observations of the thin and the thick disc (e.g. 
\citealt{shoenrich_2009, minchev_2013, kubryk_2015}). 
Work is now in progress to study the lithium evolution of the thin and 
thick discs in the framework of the model of \citet{kubryk_2015}, but lithium destruction and 
atomic diffusion in stellar interiors makes the situation more complex than for oxygen (Prantzos et al. in preparation).} 

\section{Summary}\label{conclusiooooonnnn}

Produced during the Big Bang, in stars, and in the interstellar medium, 
lithium can also be easily burnt in stellar interiors. This chemical species therefore shows a great 
interest when one tries to understand its chemical evolution in our Galaxy, 
the Milky Way.

In order to understand its chemical evolution history, 
we built a homogeneous catalogue 
of lithium abundance composed of $7\,300$ stars and based on high-resolution 
FEROS, HARPS, and UVES spectra parametrized within the AMBRE Project. We performed 
a fast automatic determination of lithium abundances by coupling a synthetic spectra grid and the 
Gauss-Newton algorithm GAUGUIN. 
It is the first time that such a massive catalogue is built 
in a very homogeneous way compared to previous studies based on few hundreds 
of stars.

The AMBRE/Li catalogue covers 
a large domain in metallicity and evolutionary stages. 
These derived lithium abundances were validated by comparison
with independent and recent literature values based on HARPS spectra.
An additional check for the Gaia benchmark stars was also performed.
From this large and unique data set, we studied the lithium  content in the Milky Way together 
with its temporal evolution.

First, based on a subsample of $2\,310$ dwarf stars, we showed that the ISM lithium 
abundance increases by $\sim1\,$dex over the domain $-1.0\le\metah\le+0.0\,$dex. This result 
is qualitatively in agreement with chemical evolution models of lithium in the Milky Way 
\citep{romano_2001, prantzos_2012}. We also find  a singular behaviour for lithium, whose
ISM abundance decreases by about $0.5\,$dex at super-solar metallicities 
($+0.0\le\metah\le+0.5\,$dex). This behaviour is not predicted by current models.

Using a robust chemical separation, based on the enrichment on $\alpha$ elements 
with respect to iron, in our sample we identified  stars as members of the thin and 
thick discs or members of a group of metal-rich $\alpha$-rich stars. We found that the thick disc 
stars suffered  a low chemical enrichment, showing lithium abundances that are rather close 
to the Spite plateau. Nevertheless, the close thin disc stars clearly show a strong 
increase of their lithium content with metallicity, probably due to 
the contribution of low-mass stars. We concluded that the thin 
and thick discs clearly show two distinct  lithium enrichment histories. 

\begin{acknowledgements}
The spectra calculations were performed with the high-performance 
computing facility MESOCENTRE, hosted by OCA. CCW would like to 
acknowledge the European Union FP7 programme through ERC grant 
number 320360 and the Leverhulme Trust through grant RPG-2012-541. 
This work was partly supported by the European Union FP7 programme 
through ERC grant number 32036. This work has made use of the 
VALD database, operated at Uppsala University, the Institute 
of Astronomy RAS in Moscow, and the University of Vienna.
\end{acknowledgements}

\bibliographystyle{aa}
\bibliography{gguiglion}

\end{document}